\documentclass[twocolumn,showpacs,aps,prd,amssymb,nofootinbib,nobibnotes,floatfix,longbibliography]{aastex631}

\usepackage{newtxtext,newtxmath}
\usepackage{ae,aecompl}
\usepackage[mathscr]{euscript}

\usepackage[T1]{fontenc}
\usepackage{hyperref}
\usepackage{amsfonts}
\usepackage{amsmath}
\usepackage{mathrsfs}
\usepackage{epsfig}
\usepackage{graphicx}
\usepackage{url}
\usepackage{hyperref}
\usepackage{float}
\usepackage{color}
\usepackage[utf8]{inputenc} 
\usepackage{epsf}
\usepackage{comment}
\usepackage{slashed}
\usepackage{footmisc}
\usepackage{lipsum}
\definecolor{linkcolor}{rgb}{0.0,0.3,0.5}
\usepackage[caption=false]{subfig}

\usepackage{booktabs}

\usepackage{color}
\definecolor{cerulean}{rgb}{0.0, 0.48, 0.65}

\definecolor{navy}{rgb}{0.2, 0.0, 1.0}

\definecolor{jungle}{rgb}{0.0, 0.5, 0.0}

\usepackage{color}

\definecolor{orange}{rgb}{1,0.5,0}
\definecolor{orangeB}{rgb}{1,0.7,0}

\usepackage{ulem}
\usepackage{units}
\usepackage{soul}
\usepackage{orcidlink}

\usepackage[mathlines]{lineno}% Enable numbering of text and display math

\begin{document}

\preprint{APS/123-QED}

\title{Little Red Dots as self-gravitating discs accreting on supermassive stars:\\ Spectral appearance and formation pathway of the progenitors to direct collapse black holes}

\author{Lorenz Zwick}
\affiliation{Niels Bohr International Academy, The Niels Bohr Institute, Blegdamsvej 17, DK-2100, Copenhagen, Denmark}
\affiliation{Center of Gravity, Niels Bohr Institute, Blegdamsvej 17, 2100 Copenhagen, Denmark.}

\author[0000-0002-3820-2404]{Christopher Tiede\,\orcidlink{0000-0002-3820-2404}}
\affiliation{Niels Bohr International Academy, The Niels Bohr Institute, Blegdamsvej 17, DK-2100, Copenhagen, Denmark}

\author{Lucio Mayer}
\affiliation{Department of Astrophysics, University of Zurich, Winterthurerstrasse 190, 8057, Zurich, Switzerland}

\shorttitle{LRD from highly accreting SMS}
\shortauthors{Zwick et al.}

\date{\today}% It is always \today, today,
             %  but any date may be explicitly specified

\begin{abstract}
\noindent
We propose an alternative physical interpretation and formation pathway for the recently discovered "little red dots" (LRDs). We model LRDs as super-massive stars (SMSs) surrounded by massive self-gravitating accretion discs (SMDs) that form as a consequence of gas-rich major galaxy mergers. The model provides an excellent match for numerous spectral features of LRDs, where the V-shape arises from the superposition of two black bodies, and Balmer line broadening is sourced by the intrinsic rotation of the SMD. No additional AGN, stellar, dust, or {broadening component is strictly} required. This results in a model with physically motivated parameters that are robust to variations in observed LRD properties. We perform MCMC fits for two representative LRD spectra, for which the full parameter posterior distributions are determined. Allowing for a compressed SMS mass-radius relation, the recovered parameters are compatible with sub-Eddington accretion in self-gravitating discs, and the recovered SMS masses of few $ 10^6$ M$_{\odot}$ imply the subsequent formation of massive black holes (BH) that squarely follow the expected BH mass--galaxy mass relation{, while also predicting a cut-off luminosity of order few $10^{44}$ erg/s in quantitative agreement with current observations}. {While matching the abundance of LRDs is challenging, the association to galaxy mergers produces a redshift distribution that reflects observations.}
\end{abstract}

% ============================================================
\section{Introduction}
\label{S:Introduction}
The presence of supermassive black holes (SMBHs) with masses exceeding $10^9$ M$_{\odot}$ at redshifts $z \gtrsim 6$ presents a fundamental challenge to our understanding of structure formation in the universe \citep[see, e.g.][]{Bolton63,Schmidt63,1964zeldovich,2011priya,2012Volonteri,2013Haiman,2019Mayer,Woods2019,2017Mazzucchelli,2020sloan,wang2021,xuheng}.  Observed less than a billion years after the Big Bang, these objects appear far too massive to have formed naturally as a consequence of Population-III stellar evolution and the subsequent Eddington limited accretion of gas \citep{Madaurees2001,Abel2002,Schneider2002,Hirano2014}.

The proposed resolutions to this enigma (excluding primordial BH scenarios, e.g. \citealt{2018Bellomo,Zhou2022,2022zhou2}) can be classified into two categories. The first involves invoking long periods of super-Eddington accretion, which can grow BHs from small initial masses ($\lesssim  10^3$ M$_{\odot}$) to the observed values at redshifts of $\gtrsim 6$. However, numerical works suggest that it is implausible to sustain such accretion rates over the required hundreds-of-millions of years due to feedback effects, black hole wandering, or a simple lack of sufficient gas supply \citep{2008alvarez,2018smith,Zhu2020,Sassano2021,2016MNRASfiacconi,2016lupi,Regan2019,2022sassano,2025husko}. 

The second category involves the formation of so called "heavy seed" BHs with masses $\gtrsim10^4$ M$_{\odot}$ through the direct collapse of primordial gas clouds. 
The most common formulation of the direct collapse scenario is via the contraction of pristine gas in dark matter halos that are illuminated by Lyman-Werner radiation from a nearby galaxy of Pop III stars \citep{agarwal,2013Latif,2014regan,2016habouzit,2016agarwal,2017johnson,2024prole}. 
Several complementary models aim to circumvent the need for external radiation by invoking a number of other likely relevant physics such as dynamical heating, turbulent inflow from streams, or global disc instabilities \citep{Shlosman:BarInBar:1989, LodatoNatarajan:2006,Colgate2003,2004koushippas,2006begelman,2008begelman,2022latif}. Common to these alternate formulations of direct collapse are extremely large gas inflows and the consequent formation of a highly accreting, hydrostatic spherical structure that will later undergo gravitational collapse. These structures have been modeled either as ``quasi-stars'' (QS) \citep{2006begelman,2025begelman} or supermassive stars \citep[SMS;][]{fowler1966,bisnovatyi1967,begelman2010,2012hosokawa,hosokawa2013,haemmerle2018a,haemmerle2018b,2023herrington} that culminate in sudden direct collapse due to the onset of catastrophic neutrino cooling or the general relativistic instability, respectively \citep{1964chandra,2009saijo,begelman2010,shibata2016a,2020lionel} .

The JWST observation of a SMBH at $z \gtrsim 10$ with an estimated mass of $\sim 10^{7.6}$ M$_{\odot}$ \citep{Natarajan:UHZ1:2024, Bogdan:UHZ1:2024} provides even greater need for the formation of intrinsically heavy black hole seeds through direct collapse. Together with other outliers \citep[e.g.][]{2020kroupa,wang2021}, these systems provide strong evidence that direct collapse should occur in the early universe; and it further suggests that direct collapse models may need to produce seeds with mass $\gtrsim 10^5$ M$_{\odot}$. Recently, from the observations of a novel class of JWST objects, colloquially, the ``Little Red Dots'', several authors have started to suggest that we may be directly observing BH seeds, as either super-Eddington accretors \citep{PacucciNarayan:LRDs:2024} or as post-direct-collapse systems themselves \citep{NaiduMatthee:2025}. Little red dots (LRDs) are selected by their compact (point-like) size and characteristic V-shaped spectra with red slopes in the rest-frame optical, but flat or blue colours into the UV \citep{Matthee+2024-LRD}. Most LRDs present broad Balmer emission lines, suggesting that they comprise a subset of massive BHs accreting like active galactic nuclei (AGN) \citep{Greene:LRDs:2024}.
However, producing the spectral V-shape is a challenge in standard accretion modeling, and LRDs are also almost entirely lacking in hallmark X-ray emission \citep{Yue:LRDs:2024, Sachhi:LRD-Xrays:2025}. Numerous authors have also suggested that standard relations to estimate the central mass (and bolometric luminosities) from the broad lines (and optical emission) seem to break down and over-predict both quantities \citep{Ananna:LRDs:2024}.

In this work, we propose an alternative physical interpretation and formation pathway for LRDs. In this picture, major mergers of galaxies with $\sim 10^9$ M$_{\odot}$ of gas trigger strong inflows that result in a compact self-gravitating disc (SMD) at sub-pc scale, with typical temperatures of $\sim 4000$ K. The SMD feeds a highly accreting supermassive star (SMS) which radiates as a hot black body ($\sim 20000$ K). Within this picture, we recover numerous puzzling features of LRD spectra and their cosmic abundance by allowing for a compressed SMS mass-radius relation.  Crucially, the recovered parameters from spectral fitting are compatible with standard angular momentum transport mechanisms and standard Eddington limited accretion.

We note that our interpretation of LRDs shares some conceptual similarities with the models recently proposed in \cite{ZhangWu:TwodiskLRD:2025}, a self-gravitating disc, and \cite{2025begelman,2025nandal}, a SMS-like structure. However, it differs in what parts of LRD spectra are associated to what structure and provides a unified treatment for the SMS and its accretion flow. Our model does not require any additional AGN, stellar wind, dust obscuration or galactic stellar component, resulting in few physically motivated free parameters that are robust to variations in observed LRD properties. Additionally, our model uniquely makes an explicit connection between LRDs and gas rich galaxy mergers, tying these novel objects with the physics of high redshift structure formation and allowing us to model the evolution of the LRD co-moving density.

The paper is structured as follows: In Sections~\ref{S:physical-model}, \ref{sec:sms} and \ref{sec:smd} we detail our model for a highly accreting SMS being fed by a SMD, noting that it naturally predicts the star and disc temperatures to match the V-shape of LRDs with very weak dependencies on physical parameters. In Section~\ref{S:spectra-fitting} we apply our model to a representative set of LRD spectra and comment on the inferred SMS and SMD properties. In Section~\ref{S:Dicsussion} we thoroughly discuss the spectral properties of the model in the context of LRD observations, and compare the redshift distribution of observed LRDs with estimates from galaxy mergers. Finally, we present a brief summary and some concluding remarks in Section~\ref{S:conclusion}.

% ============================================================
\section{SMS/SMD model for LRDs}
\label{S:physical-model}

\begin{figure}
    \centering
    \includegraphics[width=1\linewidth]{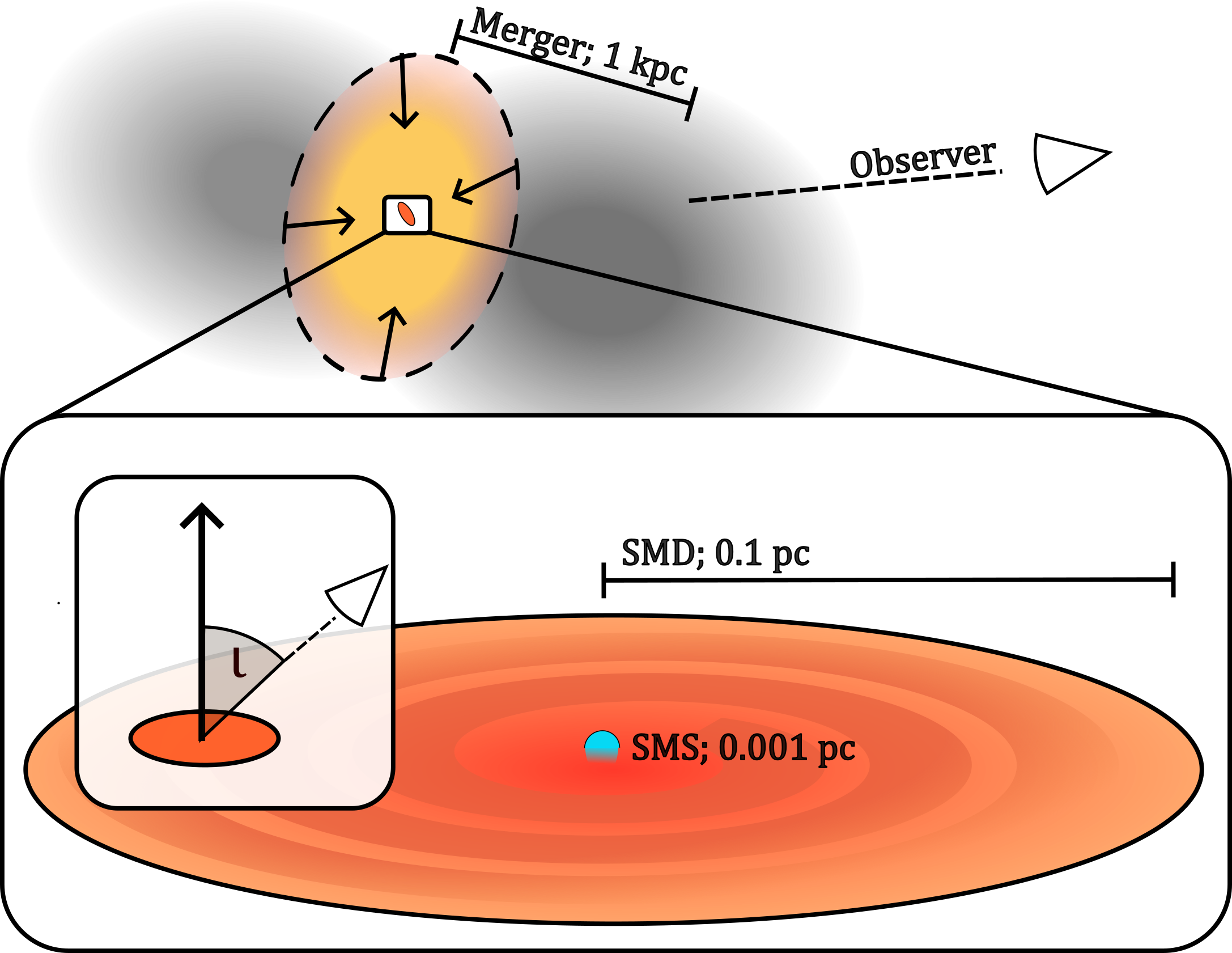}
    \caption{Simple cartoon of the assembly of a supermassive disc (SMD) and supermassive star (SMS) system that appear as little red dots (LRD). Major mergers of galaxies with at least $\sim 10^8$ M$_{\odot}$ of gas trigger strong inflows that result in a compact self-gravitating disc at sub-pc scale, with typical temperatures of $\sim 4000$ K. The disc feeds an accreting SMS which radiates as a hot black body ($\sim 20000$ K).}
    \label{fig:illustration}
\end{figure}

% ------------------------------------------------------------
Highly accreting SMSs in particular provide an extremely interesting precursor stage to direct collapse BHs, as they can reach masses exceeding $\sim 10^{6}$ M$_\odot$ before collapse. This threshold mass can be even larger if other forms of support such as rotation or magnetic fields are present \citep{shibata2016a,chon2018,2021lionel,2024kiyuna,2025lionel}. Crucially, this makes possible the creation of very massive BH seeds that retain the majority of the stellar mass \citep{baumgarte1999b,2009saijo,2013reisswig,Woods2019,2021woods} and ensures that the system can endure large accretion rates over long timescales.
While SMSs remain theoretical objects, both analytical studies and numerical simulations have demonstrated that SMS-like objects, accompanied by a large scale self-gravitating accretion disc, form as a consequence of extreme gas accretion rates induced by atomic cooling halos \citep{2020regan, 2021woods} or the major mergers of gas rich galaxies \citep{Mayer2010,Mayer2015,2014bonoli,2019Mayer,2023MNRAS.518.2076Z,2024ApJ...961...76M,2025lionel}. These accretion rates can be sustained down to sub-pc scales and ultimately feed a central SMS by forming a centrifugally supported disc at the boundary of gravitational instability.

Our model is based on this physical picture, and is visualised in Fig.~\ref{fig:illustration}. 
The bulk of the emission consists of a superposition of two black bodies (BB) radiating according to Planck's law:
\begin{align}
    B_{\lambda} = \frac{2 hc^2}{\lambda^5}\frac{1}{\exp\left(\frac{h c}{k_{\rm B}T}\right)-1}
\end{align}
where $B_{\lambda}$ is the specific intensity at a wavelength $\lambda$, $h$ is Planck's constant, $k_{\rm B}$ is Boltzmann's constant, $c$ is the speed of light and $T$ the BB temperature. Note that to account for redshift, the wavelengths are multiplied by the factor $\lambda_{\rm z}=(1+z)\lambda_{\rm rest}$, while the intensity is decreased by a factor $B_{\lambda}^{\rm z} = (1+z)^{-3}B_{\lambda}^{\rm rest}$. The peak  wavelength of the UV component of LRD spectra is $\lambda_{\rm hot}^{\rm peak} \approx 0.15$ $\mu$m in the rest frame \citep[see e.g. the spectra in][]{Matthee+2024-LRD}. 
From Wien's Law, this corresponds to a characteristic temperature:
\begin{align}
    T_{\rm hot} &\sim 1.5 \times 10^4 \, {\rm K} \text{  to  }  2 \times 10^4\, {\rm K}.
\end{align}
Similarly, the red component of the characteristic LRD spectrum often continues beyond 1 $\mu$m, implying that the colder component must have a temperature of:
\begin{align}
T_{\rm cold}  \sim2000 \, {\rm K} \text{  to  }  5000\, {\rm K}.
\end{align}
These basic features are mentioned in several works \citep[see e.g.][]{2024setton,ZhangWu:TwodiskLRD:2025, Inayoshi:bLRDs:2025, liuquataert2025}, noting that the LRD spectral-V is well approximated by the superposition of two blackbodies with these characteristic temperatures. Given the difference in temperature and the similarity of the flux in the hotter and the colder portions of typical LRD spectra, we can also deduce that the hot BB component must have an emitting area that is much smaller than the cold component, roughly by a factor $(T_{\rm cold}/T_{\rm hot})^2\sim 100$.

In our model, the hot component consists of a highly accreting SMS, {which requires two parameters consisting of the} {stellar} radius $R_{\rm S}$ and the effective temperature $T_{\rm S}$. The star is accreting from a surrounding, cooler{, isothermal} accretion disc {which is specified by a} characteristic radius $R_{\rm D}$, temperature $T_{\rm D}$, and  scale-height $h_{\rm D}$. 
{Here, the isothermal condition is a product of the disc's marginal gravitational stability (discussed in Section~\ref{sec:smd}), and not an additional assumption.}

The rest frame BB flux $F_{\lambda}^{\rm S}$ seen at a luminosity distance $D_{\rm L}$ from the SMS is:
\begin{align}
    F_{\lambda}^{\rm S} = \pi\frac{R_{\rm S}^2}{D_{\rm L}^2} B_{\lambda}(T_{\rm S}),
\end{align}
and does not depend on the viewing angle (valid under the assumption of a thin accretion disc, which we will discuss in more detail in sections \ref{sec:smd} and \ref{S:spectra-fitting}). In contrast, the disc's BB flux $F_{\lambda}^{\rm D}$ depends strongly on viewing angle. We define the projection angle $\iota$ to range between 0 and $\pi/2$, where $\iota=0$ represents a fully face on disc, while $\iota=90^{\circ}$ represents an edge on disc:
\begin{align}
    F_{\lambda}^{\rm D} = \pi\frac{4\pi R_{\rm D}^2\cos(\iota) + 4H_{\rm D} R_{\rm D}\sin(\iota)}{4\pi D_{\rm L}^2} B_{\lambda}(T_{\rm D}).
\end{align}
\begin{figure}
    \centering
    \includegraphics[width=1\linewidth]{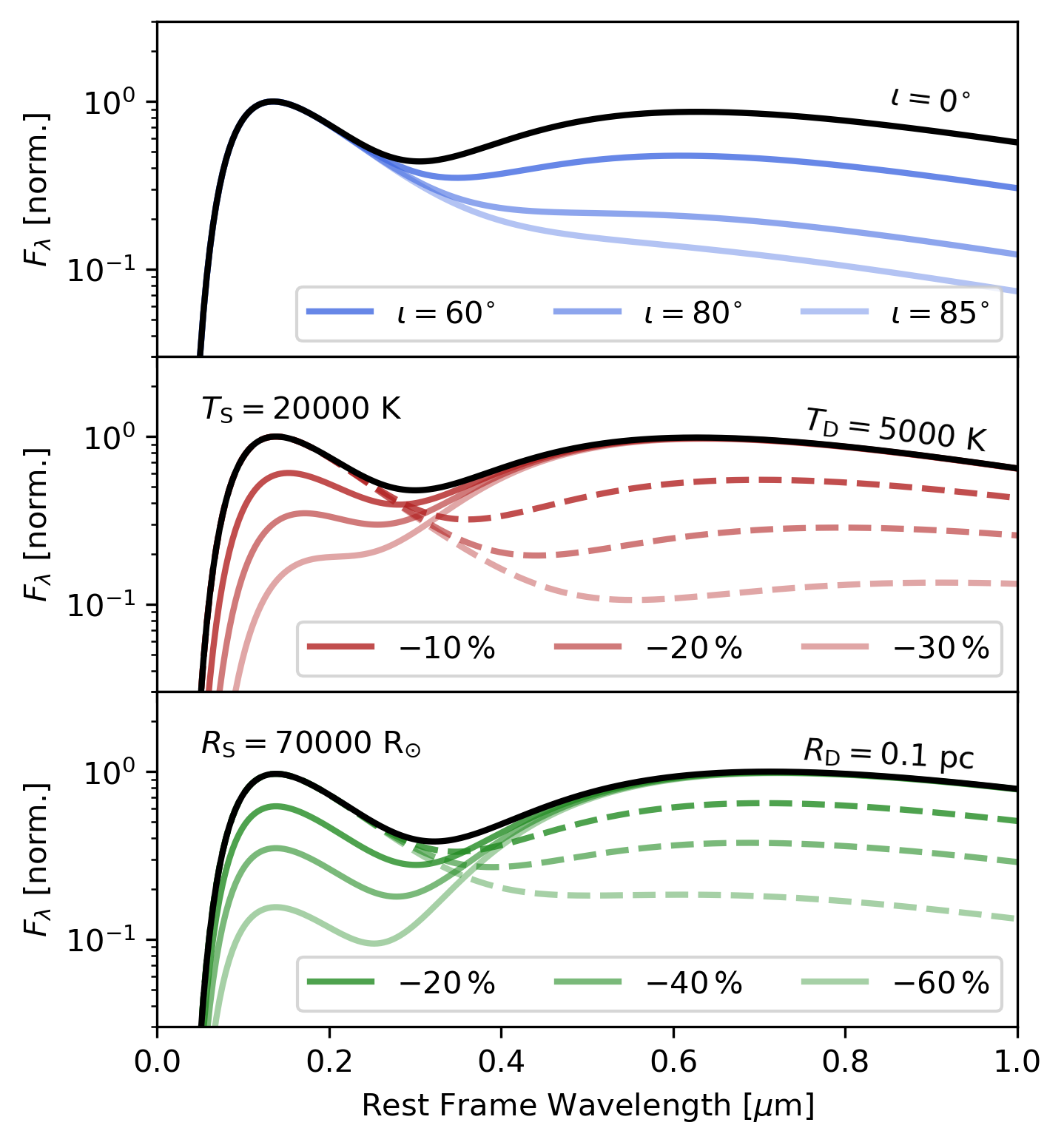}
    \caption{Basic spectra of the composite SMS/SMD model for different system inclinations (top panel), component temperatures (middle panel) and component sizes (lower panel). The chosen reference parameters are representative for the best fits of typical LRD spectra (see section \ref{S:spectra-fitting}). We highlight the different behaviours of the model by changing a single parameter at a time by the amount shown in the labels, with respect to a baseline choice represented by the black lines. In the bottom two panels, solid-lines indicate the hot SMS component and dashed-lines the cool SMD. Note that the inclination, mass and temperature of the disc become fully determined by spectral fitting once additional physical constraints are adopted (see section \ref{S:spectra-fitting}).}
    \label{fig:projections}
\end{figure}

Lastly, the majority of LRDs present with broadened H$_{\alpha}$ and H$_{\beta}$ lines which are typically attributed to Keplerian gas clouds rotating about a central SMBH, known as broad-line regions \citep{Antonucci:1993fe}. 
{Recent works have suggested that Balmer lines in LRDs may alternatively be scatter-broadened with only a more narrow component attributed to the line-of-sight motion of emitting material \citep{Rusakov:Nature:2026}}.
In our model, {we restrict the line broadening mechanism to solely that caused by the rotation of the accretion disc under its own self-gravity, though we will mention in section \ref{sec:rates} how relaxing this assumption could alleviate some astrophysical tensions regarding formation rates.}
{Thus,} the line-broadening velocity is given by the disk's circular velocity:
\begin{align}
    v_{\rm D}^{\rm rot}\sin(\iota) \equiv  \sqrt{\frac{G M_{\rm D}}{R_{\rm D}}}\sin(\iota),
\end{align}
where $M_{\rm D}$ is the total mass of the disc and we projected the disc's rotation along the line of sight. Reaching rotational velocities of thousands of km/s requires discs that are more massive and compact than atomic cooling halos \citep{2013prieto,2020regan,2024prole}. The required typical scales are:
\begin{align}
    v_{\rm D}^{\rm rot} \approx 2000\, \left(\frac{0.1\, {\rm pc}}{R_{\rm D}} \cdot \frac{M_{\rm D}}{10^8\, {\rm M}_{\odot}}\right)^{1/2} {\rm km/s}\,.
\end{align}
The formation of such ``super massive'' discs (SMDs) has been verified to occur in high-resolution numerical studies of major galaxy mergers {with stellar and gas masses of $\sim 10^{10}$ M$_{\odot}$} from both individual merger simulations \citep{Mayer2010,Mayer2015} as well as full cosmological zoom-in simulations \citep{2024ApJ...961...76M} because of the rapid removal of gas angular momentum in shocks or from galactic tidal forcing \citep[see also][]{1996mihos,2012lambas}. More broadly, SMDs or similar forms of accretion flows are motivated by the immense gas requirements necessary to fuel any model of BH growth and explain the high masses of quasars. We will discuss the expected properties of SMDs in more detail in section \ref{sec:smd}, {and discuss the plausibility of matching LRD abundances in section \ref{sec:rates}}.

To summarise, our model for LRDs consist of a smaller, hotter accreting SMS surrounded by an extended, {cooler}, self-gravitating SMD. {The core model parameters are the SMS radius and temperature, the SMD radius and temperature, the SMD aspect ratio and rotational velocity, and the system line of sight inclination. These are summarized in Table \ref{tab:params1}} and their effect on the basic spectral model is shown in Fig. \ref{fig:projections}. As we will see in the following sections, these parameters {must be supplied with a set of} physical relations and constraints in order to determine the SMS and SMD mass, stability, lifetime and other properties by fitting the resulting model spectra to observations of LRDs. {After the physical relations have been appliead, it is more convenient to use a different set parameters in the MCMC fitting procedure (shown instead in Table \ref{tab:paramsfc}).}
\begin{table}[]
    \centering
    \begin{tabular}{c|c}
        $R_{\rm S}$ & SMS radius \\
        $T_{\rm S}$ & SMS temperature \\
        $R_{\rm D}$ & SMD radius \\
        $T_{\rm D}$ & SMD temperature \\
        $h_{\rm D}$ & SMD aspect ratio\\
        $v_{\rm D}^{\rm rot}$ & SMD rotational velocity\\
        $\iota$ & System line of sight inclination
    \end{tabular}
    \caption{The {basic BB parameters of the} SMS/SMD spectral model, which are related to physical properties of SMSs and self-gravitating discs as detailed in sections \ref{sec:sms} and \ref{sec:smd}. These are to be understood as characteristic values that describe the general features of the spectrum. The parameters are recovered by fitting a two component BB model to the observed spectra, associating the broadening of the hydrogen lines to the rotation of the SMD and enforcing several physical constraints for both the SMS and the SMD.}
    \label{tab:params1}
\end{table}

% ------------------------------------------------------------
\section{Accreting supermassive stars}
\label{sec:sms}

\subsection{SMS accretion and temperature}
{In realistic astrophysical scenarios, the very formation of supermassive stars (SMS) relies on strong accretion flows that provide the required mass budget in less than a nuclear burning timescale \citep[see e.g.][]{hosokawa2013}. The structure and effective temperature of SMSs are strongly influenced by whether ongoing accretion dominates the entropy profile of the star:}
{SMS of masses $\gtrsim10^4$ M$_{\odot}$ typically behave like bloated proto-stars for rates above $\sim 0.1$~M$_{\odot}$~yr$^{-1}$ \citep{2012hosokawa,hosokawa2013,haemmerle2018a,haemmerle2019c}, and evolve with a roughly constant effective temperature of $\sim 5000$ K. In this work, we invoke the presence of a SMS to account for the temperatures of $\sim 10^4 \,\mathrm{K}$ inferred for the hot component of the LRD spectra. To reconcile this aspect we will postulate a phenomenological class of SMS that have a more compact mass-radius relation. We detail the features of such compactified SMS below, and discuss their plausibility in section \ref{sec:constraints}}. 

{First let us review the typical mass-radius relation used for bloated accreting SMS. Numerical simulations for accretion rates of order $\gtrsim 0.01$ M$_{\odot}$ yr$^{-1}$ have found the radius of accreting SMSs to be well approximated by \citep{hosokawa2013}:
\begin{align}
    \label{eq:hosokawa}
    R_{\rm S}^{\rm Hos}\approx 2.6\times10^4\, {\rm R}_{\odot}\left(\frac{M_{\rm S}}{10^4{\rm M}_{\odot}} \right)^{1/2},
\end{align}
up to masses of several $ 10^4$ M$_{\odot}$, and this relation is often extrapolated to higher masses \citep[see e.g.][]{lionel2021}. This corresponds to the outer envelopes of SMSs being bound by the Hayashi limit, which marks the onset of global convective instabilities \citep{1961hayashi}. Bloated accreting SMS are by definition in a regime where the thermal time exceeds the characteristic accretion time, implying that the emergent luminosity is predominantly set by accretion. To estimate the accretion luminosity,  consider that the kinetic energy of gas in the accretion flow is thermalised via shocks and other dissipative processes \citep{2002frank} at the stellar surface. The typical velocity of a fluid element close to the SMS is dominated by the star's gravitational potential, and is therefore of order of the circular velocity:
\begin{align}
    v^{\rm c}_{\rm S} = \sqrt{\frac{G M_{\rm S}}{R_{\rm S}}},
\end{align}
and the associated accretion power is:
\begin{align}
   P_{\rm acc} =  \frac{\eta_{\rm acc}}{2} \dot{M}_{\rm S} v_{\rm c}^2,
\end{align}
where the efficiency of the energy conversion is described via a dimensionless parameter $\eta_{\rm acc}$, and $\dot{M}_{\rm S}$ is the accretion rate onto the SMS. To obtain the associated SMS temperature, we equate the accretion power to the radiated BB power $P_{\rm BB}= 4\pi R_{\rm S}^2\sigma_{\rm B}T_{\rm S}^4$ of the SMS, yielding a relation between the star's temperature, mass, radius and accretion rate:
\begin{align}
    T_{\rm S}^4 = \frac{\eta_{\rm acc}}{8\pi}\frac{G  M_{\rm S} \dot{M}_{\rm S}}{\sigma_{\rm B} R_{\rm S}^3},
\end{align}
Evaluating for some typical SMS parameters gives:
\begin{align}
\label{eq:T_S}
    T_{\rm S}&\approx 6.11\times10^3\,  \left(\eta_{\rm acc }\frac{M_{\rm S}}{10^5\, {\rm M}_{\odot}} \cdot \frac{\dot{M}_{\rm S}}{10^2\,{\rm M}_{\odot}\, {\rm yr}^{-1}}\right)^{1/4}\nonumber \\ &\times \left(\frac{R_{\rm S}}{5\times 10^4\,R_{\odot}}\right)^{-3/4}{\rm K}.
\end{align}
and we in fact recover typicaly quoted SMS temperatures of $\sim 5000$ K.}

{We now make the assumption that more compact SMS can exist even in a high accretion regime (see section \ref{sec:constraints} for a thorough discussion). We parametrise the SMS mass radius relation phenomenologically by introducing a compression factor $f_{\rm c}<1$, such that $R_{\rm S} = f_c R_{\rm S}^{\rm Hos}$ and:
\begin{align}
     \frac{M_{\rm S}}{\rm{M}_{\odot}} &= f_{\rm{c}}^{-2}\,\left(\frac{ R_{\rm S}}{260\,{\rm R}_{\odot}} \right)^2 \ .
\end{align}
This provides an estimate of the SMS mass given a recovered value for the SMS radius from the hot BB component of our model spectrum, under the assumption of a phenomenological modification to the mass-radius relation for bloated SMS dominated by accretion}.

The SMS mass radius relation also sets a maximum allowed accretion rate onto the surface of the SMS that is consistent with purely gravitational physics, {disregarding for the moment the Eddington limit}. This gravitational accretion limit has been derived in \cite{lionel2021} following the  mass-radius relation of Eq. \ref{eq:hosokawa}. We report it here with a simple modification to account for a possible compression:
\begin{align}
    \label{eq:accrate}
    \dot{M}_{\rm S}^{\rm max}\approx   2\times 10^4\, f_{\rm c}^{-3/2}\,\left( \frac{M_{\rm S}}{10^4\,\rm{M}_{\odot}}\right)^{3/4}\,{\rm M}_{\odot}\, {\rm yr}^{-1},
\end{align}
where the scaling of $f_{\rm c}^{-3/2}$ arises from the simple dependence of the limit with the gravitational compactness of the accretor. We parametrise the actual accretion rate on the SMS as a fraction $f_{\rm acc}<1$ of the gravitational maximally allowed rate, i.e. $\dot M_{\rm S} = f_{\rm acc}\, \dot{M}_{\rm S}^{\rm max}$. Given this parametrisation, we can use the mass radius relation and the accretion rate to express the effective BB temperature of a SMS simply as a function of its mass and a few dimensionless numbers:
\begin{align}
    T_{\rm S}^4 = \frac{\eta_{\rm acc} f_{\rm acc} f_{\rm c}^{-9/2}}{8\pi}\frac{G  M_{\rm S} (20\, {\rm M_{\odot}\, yr^{-1}})}{\sigma_{\rm B} (260 \, {\rm R}_{\odot})^3 } \left( \frac{M_{\rm S}}{M_{\odot}}\right)^{-3/4} .
\end{align}
Evaluating for some typical parameters gives:
\begin{align}
    \label{eq:typ_temp_sms}
    T_{\rm S}\approx  2.4 \times 10^4 \, \eta_{\rm acc}^{1/4} f_{\rm acc}^{1/4} f_{\rm c}^{-9/8}  \left(\frac{M_{\rm S}}{10^5\, {\rm M}_{\odot}}\right)^{1/16}\,{\rm K}.
\end{align}
For efficiencies $\eta_{\rm acc}\lesssim 1$, we see that the temperature of tens of thousands of Kelvin required to explain the UV emission peak in LRD spectra is, in principle, entirely consistent with {gravitationally} allowed accretion rates of $f_{\rm acc}<1$. {We also note how introducing the phenomenological compression factor $f_{\rm c}$ greatly reduces the required accretion rate necessary to achieve a given temperature.} {With Eq. \ref{eq:typ_temp_sms}, we have therefore shown that highly accreting SMS can also provide a physical origin for the hot part of LRD spectra, if:
\begin{itemize}
    \item They accrete at extreme, though not gravitationally impossible, rates.
    \item They are compressed with respect to the Hayashi limit (Eq. \ref{eq:hosokawa}.)
\end{itemize} 
As mentioned above, we will specifically focus on the latter case, and discuss its plausibility thoroughly in section \ref{sec:constraints} and \ref{sec:fc}.}

{Up to this point, we have not enforced} sub-Eddington accretion explicitly in our model. For accretion by a thin disk, the Eddington rate is
\citep{1921eddington,2002frank}:
\begin{align}
    \label{eq:eddi}
     \dot{M}_{\rm S}^{\rm Edd}  = \frac{8 \pi m_{\rm p}   c}{\eta_{\rm acc}\sigma_{\rm T}} R_{\rm S} \approx 104.5\,\left(\frac{R_{\rm S}}{5\times 10^4 \, {\rm R}_{\odot}} \right)\, {\rm M}_{\odot}\, {\rm yr}^{-1}.
\end{align}
{Instead, we will apply it after the spectral} fitting {as a consistency check. Here, we note that, by definition, accretion at the Eddington limit delineates the transition between an accretion dominated proto-star and a radiation pressure dominated star powered by fusion processes. In our spectral fitting,} we will find fits that do not require exceeding the Eddington limit {by invoking $f_{\rm c} \lesssim 0.1$, a value that is also suggested by several additional considerations discussed in section \ref{sec:constraints} and \ref{sec:fc}}.

{Crucially however, as long as the accretion rate through the disk is at least comparable to the Eddington rate at the stellar surface, the accretion luminosity necessarily constitutes a substantial fraction of the total stellar luminosity. In this regime, the effective temperature given by Eq.~\ref{eq:typ_temp_sms} remains a good approximation, largely independent of the detailed internal structure of the SMS. We use this fact to justify neglecting to model the specifics of the SMS interior in the context of a broad, physically motivated model for LRD spectra.} Finally, note that the Eddington limit on a SMS is much larger than the equivalent limit on a BH with comparable mass, due to accretion occurring at larger scales in a much shallower region of the gravitation potential.

\begin{table}[]
    \centering
    \begin{tabular}{c|c}
        $R_{\rm S}$ & SMS radius \\
        $T_{\rm S}$ & SMS temperaturee \\
        $R_{\rm D}$ & SMD radius \\
        $T_{\rm D}$ & SMD temperature \\
        $h_{\rm D}$ & SMD aspect ratio\\
        $n_{\rm D}$ & SMD number density\\
        $\iota$ & System line of sight inclination \\ \hline
        \newline \newline 
        $f_{\rm c}$ & SMS compression factor \\
    \end{tabular}
    \caption{{The parameters of the SMS/SMD model after all relevant constraints have been applied. The MCMC spectral fitting is performed using these parameters, resulting in entirely determined system properties up to the choice of $f_{\rm c}$. We highlight the parameter $f_{\rm c}$, as it is a phenomenological modification the a SMS mass-radius relation that follows the Hayashi limit.}}
    \label{tab:paramsMCMC}
\end{table}

% ------------------------------------------------------------
\subsection{{Constraints for compressed and accreting SMS}}
\label{sec:constraints}
Collecting Eqs. \ref{eq:T_S} and \ref{eq:accrate} we find that a measurement of SMS temperature and SMS radius from the spectral model ($T_{\rm S}$ and $R_{\rm S}$) gives a  constraint on the dimensionless parameters $\eta_{\rm acc}$, $f_{\rm acc}$ and $f_{\rm{c}}$,
\begin{align}
     \eta_{\rm acc} f_{\rm acc} f_{\rm c}^{-9/2}  \approx \left(\frac{M_{\rm S}}{10^5\, {\rm M}_{\odot}}\right)^{-1/4}\left(\frac{T_{\rm S}}{2.4\times 10^4\,{\rm K}}\right)^4.
\end{align}
Replacing the SMS mass with the SMS radius and rescaling we find:
\begin{align}
\label{eq:SMS_constraint}
    \eta_{\rm acc} \, f_{\rm acc} \, f_{\rm{c}}^{-5}  = 0.4\times \left(\frac{T_{\rm S}}{1.8 \times 10^4\, {\rm K}} \right)^4\left(\frac{R_{\rm S}}{5 \times 10^4\, \rm{R}_{\odot}} \right)^{-1/2},
\end{align}
where we chose parameters that roughly match the spectra analysed in section \ref{S:spectra-fitting}. Eq. \ref{eq:SMS_constraint} must be valid for the recovered $T_{\rm S}$ and $R_{\rm S}$ to represent a physical SMS. Here we give some rough estimates for the values of $\eta_{\rm acc}$, $ f_{\rm acc}$, and in particular discuss the possible range of $f_{\rm{c}}$.

From energy conservation we must have $\eta_{\rm acc} \leq 1$. Here we assume values close to unity ($\eta_{\rm acc} =1$), which correspond to full thermalisation of the accretion flow, {which must occur due to the SMS being a structure in hydrostatic equilibrium}. By definition, $f_{\rm acc} <  1$ and we can express the following constraint:
\begin{align}
    f_{\rm acc}  = 0.4 f_{\rm c}^5\,  \left(\frac{T_{\rm S}}{1.8\times 10^4\, {\rm K}} \right)^4\left(\frac{R_{\rm S}}{5 \times 10^4 \, \rm{R}_{\odot}} \right)^{-1/2},
\end{align}
where we used Eq. \ref{eq:accrate}. The actual accretion rate scales instead as:
\begin{align}
    \dot{M}_{\rm S}\approx    2\times 10^4 f_{\rm c}^2 \,\left(\frac{T_{\rm S}}{1.8\times 10^4\, {\rm K}} \right)^4\left(\frac{\rm{R}_{\rm S}}{5 \times 10^4\,R_{\odot}} \right)\,{\rm M}_{\odot}\, {\rm yr}^{-1}.
\end{align}
For convenience, we report it here also in terms of the SMS mass:
\begin{align}
    \dot{M}_{\rm S}\approx   10^4 f_{\rm c}^3  \,\left(\frac{T_{\rm S}}{1.8\times 10^4\, {\rm K}} \right)^4\left(\frac{\rm{M}_{\rm S}} {10^{4}\,M_{\odot}} \right)^{1/2} \,{\rm M}_{\odot}\, {\rm yr}^{-1}.
\end{align}
Note the strong dependence of the accretion rate (and fraction of the maximum rate) on the compression factor $f_{\rm c}$, to which we will return in section \ref{S:Dicsussion}.

Estimating a plausible range for the compression factor is subtle since deviations from the mass radius relation of Eq. \ref{eq:hosokawa} can have several causes. As stated in \citet{2010hosokawa,2012hosokawa,hosokawa2013}, SMS mass radius relations should be revised to account for the effect of ram pressure. In our specific case, we are in a regime where the temperature (and thus, entropy) at the surface of the star is dominated by the thermalisation of accreting gas, so we would expect compression. Furthermore, it is possible for the SMS structure {to evolve away from the Hayashi\footnote{The latter is simply a maximum theoretical value and relaxed massive stars over a wide range of mass decades typically are more compact and hotter \citep{Stahler:1980fx,Stahler:1980jl,Stahler:1981cq,Kippenhahn:1994tm,haemmerle2018a}.} limit, even}
under extreme accretion rates $\gtrsim 10^2$ M$_{\odot}$ yr$^{-1}$. In particular, one work attempting to push the accretion rates above $\sim 10^2$ M$_{\odot}$ yr$^{-1}$ found a clear, sudden increase in effective temperature at roughly constant luminosity for SMS with masses larger than $10^5$ M$_{\odot}$, implying a reduction in radius of order $\sim 10$ \citep[see the evolutionary tracks in][]{2024A&A...689A.351N,2025nandal}. Additionally, the only work that has modeled the formation of a SMS directly from angular momentum loss mechanisms in SMDs finds the central star-like object to be more compact than Eq. \ref{eq:hosokawa} by factors of 1 to 100, strongly depending on the density profile of the disc \citep{2023MNRAS.518.2076Z}. {In general, we note} that SMS have only been simulated robustly for accretion rates of up to a few $\rm{M}_{\odot}$ yr$^{-1}$, and up to masses of $\lesssim 10^5$ M$_{\odot}$. We note also that the definition of the stellar surface itself is subtle in a composite model of a stellar component plus a disc, {especially in the case of a disk accretion that can penetrate layers of the star \citep{2010hosokawa}}. In our particular case, $R_{\rm S}$ is best interpreted as the radius at which the majority of the accreting gas has thermalised and at which the bulk of the thermal radiation can escape the system. This may vary from typical definitions {for an isolated star} set by e.g. vanishing pressure or optical depth of order unity \citep{Kippenhahn:1994tm}.

Because of these considerations, we opt to simply interpret $f_{\rm c}$ as a phenomenological modification to the typical SMS mass radius relation (Eq. \ref{eq:hosokawa}), which is derived under the specific conditions detailed in \cite{hosokawa2013}. Therefore, we allow the parameter $f_{\rm c}$ to vary between 0.01 and 1. {This interval should be regarded as a nominal range. It is motivated primarily by the factor of $\sim 10$ radius contraction observed along the evolutionary tracks of \cite{2024A&A...689A.351N} for $\dot M \gtrsim 10^2$ M$_{\odot}$, as well as by the range of radii computed for SMD hydrostatic cores in \cite{2023MNRAS.518.2076Z}, which depend on SMD properties. It is also compatible with the typical sizes of SMS that are not accretion dominated \citep{haemmerle2018a}, as discussed in section \ref{sec:fc}} As we will argue in section \ref{sec:fc} and \ref{S:Dicsussion}, additional constraints can be leveraged to estimate a value for $f_{\rm c}$ in the context of applying the SMS/SMD model to LRDs. {We will find that the model requires $f_{\rm c}\lesssim 0.4$ to be entirely self-consistent with the thin disc approximation, and that selecting a value of $f_{\rm c}\sim 0.1$ provides a remarkable match with several observed properties of LRDs, while also assuring near-Eddington accretion rates.}

\subsection{GR instability}
The end of the life of a highly accreting SMS is triggered by the general relativistic instability \cite{1964chandra}. The onset of the GR instability for highly accreting SMS has been studied in detail in several works, consistently finding a limit of $\sim 10^6$ M$_{\odot}$ for non-rotating SMS through a wide range of accretion rates \citep{2024A&A...689A.351N,2025lionel}. However, other forms of support such as rotation or magnetic fields can significantly postpone the collapse \citep{fowler1966,bisnovatyi1967,baumgarte1999b,shibata2016a,lionel2021,2024A&A...689A.351N} and allow for order of magnitude larger SMS masses. {In particular, \cite{2021lionel} shows that SMS rotating at only $1\%$ of the critical velocity do not collapse until they exceed $10^7$ to $10^8$ M$_{\odot}$.} When accretion onto the SMS is provided by a self-gravitating disc, we do expect rotational support to play an important role in delaying the onset of the GR instability. {We note that none of the aforementioned works consider SMS with a phenomenological compression factor. In polytropic stars, the GR instability arises as a competition between the critical adiabatic index $\Gamma_{\rm crit}$ and the SMS compactness \citep{1964chandra}:}
\begin{align}
    \Gamma_{\rm crit} \approx \frac{4}{3} + K\frac{GM_{\rm S}}{R_{\rm S}c^2},
\end{align}
{where $K$ is a coefficient of order unity and the result is given at first post-Newtonian order. While a compressed star becomes more gravitationally compact, its density $\rho_{\rm S}$ also increases, shifting the balance between radiation and gas pressure. The adiabatic index of a mixture of gas and radiation pressure is:}
\begin{align}
\label{eq:gamma}
    \Gamma &\approx \frac{4}{3} + \frac{\beta}{6} + \mathcal{O}(\beta^2)\\
    \beta &\approx \frac{P_{\rm gas}}{P_{\rm rad}} \propto \frac{\rho_{\rm S}}{T_{\rm S}^3}
\end{align}
{From this, we can compare the stability of typical SMS against those that follow a more compact mass radius relation. To match the hot part of LRD spectra, we are considering SMSs with effective temperatures up to four times higher than the canonical SMS value of $\sim$5000 K, while being a factor $f_{\rm c}$ more compact. Since the density goes as $f_{\rm c}^{-3}$, the scaling of Eq. \ref{eq:gamma} suggests that the instability onset is delayed for $f_{\rm c} \lesssim 0.25$ with respect to a non-compressed SMS, for these specific higher effective temperatures. Note however, that the GR instability ultimately depends on the full internal structure of the SMS, and that this argument is only qualitative. Nevertheless, we are in a scenario where rotation is likely present, and the increased compactness may further delay the onset of the GR instability. Since these effects have not yet been explored with stellar evolution codes, we remain agnostic about the precise onset of the GR instability, while highlighting that several arguments point toward collapse occurring at masses above $\sim 10^6$ M$_\odot$.}

% ------------------------------------------------------------
\section{Massive self-gravitating discs}
\label{sec:smd}

\subsection{Disc radius, scale-height and temperature}
{We assume that the accretion disk is a large scale, marginally stable self-gravitating disk, where mass accretion is driven by the stresses associated with the gravito-turbulent motion of the gas.}
The formation of such SMDs has been analysed in detail in several studies with high-resolution multi-scale simulations of high-z galaxy mergers, 
including from cosmological initial conditions
\citep{Mayer2010,Mayer2015,2014bonoli,2019Mayer,2024ApJ...961...76M}. In the simulations, efficient dissipation of the merger's orbital kinetic energy via shocks and tidal torques leads to the formation of a compact SMD. The size of the disc is typically of order a pc and in the range $10^8-10^9 M_{\odot}$. This corresponds to roughly 10$\%$ of the total pre-merger gas in the galaxies. The SMDs that form are self-gravitating, with Toomre parameters in the range $1-2$ such that gravitational instabilities are expected to drive turbulence in the disc. The gravito-turbulence gives rise to super-thermal velocity dispersions in the range $10 - 20 \%$ of the rotational velocity, resulting in disc aspect ratios $\sim 0.1$, exceeding those typically associated to such cold discs (e.g. $\sim 3000-6000$ K). Due to its high density, the SMD is optically
thick with photon diffusion times exceeding the local orbital time, preventing the disc from fragmenting---except in the outermost low-density regions (see discussion in \citealt{Mayer2015,2023MNRAS.518.2076Z}). Importantly, however, such simulations have only been carried out for highly massive galaxy mergers with a stellar mass of $\sim 10^{10}$ M$_{\odot}$, {and a gas-to-star-mass ratio of $\sim 1$}. Here we assume that the general characteristics of SMD formation can be scaled to major galaxy mergers of smaller total mass. In short, we preserve that approximately 10$\%$ of the gas mass {in a major merger} participates in the inflow, ultimately settling into a compact, self-gravitating disc of a similar temperature range with scale-heights supported by underlying turbulence. {We will further discuss the plausibility of this assumption and the consequences for LRD abundances in section \ref{sec:rates}.}

We characterize the disc geometry by a radius $R_{\rm D}$ and a scale-height $H_{\rm D}$ with typical aspect-ratio $h_{\rm D}$. At the largest scales:
\begin{align}
    H_{\rm D} = h_{\rm D}\, R_{\rm D},
\end{align}
and we enforce $h_{\rm D}<1$ to preserve a thin geometry.

\subsection{Toomre stability}
The criterion for Toomre stability determines whether a disc is stabilised against fragmentation by its own shear \citep{toomre}. It is given by:
\begin{align}
    \mathcal{S}\equiv\frac{c_{\rm s} \kappa}{2\pi G \rho H_{\rm D}} > 1,
\end{align}
where $\kappa$ is the epicyclic frequency, $c_{\rm s}$ the speed of sound and $\rho$ the gas density. The epicyclic frequency is computed as: 
\begin{align}
    \kappa^2 = \frac{2 \Omega}{R} \frac{{\rm d}}{{\rm d}R}(R^2\Omega).
\end{align}
While formally this criterion is dependent on the disc's radial profile, we apply it here to estimate whether the disc as a whole can be sustained against global instabilities for a sufficient timescale. The bulk rotational frequency of the disc is $\Omega_{\rm D}=\sqrt{R_{\rm D}/(G M_{\rm D})}$, such that:
\begin{align}
    \kappa^2 \approx 2 \frac{R_{\rm D}}{G M_{\rm D}}.
\end{align}
Expressing the stability criterion in terms of our system variables gives:
\begin{align}
   \mathcal{S} &\approx 1.5 \times \left( \frac{h_{\rm D}}{10^{-2}} \cdot \frac{n_{\rm D}}{10^{14}\, {\rm cm}^{-3}} \right)^{-3/2} \nonumber \\ 
   &\times \left( \frac{R_{\rm D}}{0.1\, {\rm pc}} \right)^{-2}  \left( \frac{T_{\rm D}}{4300\, {\rm K}} \right)^{1/2},
\end{align}
where we assumed a monoatomic gas of pure hydrogen. To enforce this physical constraint, we will select disc solutions that have a nominal Toomre parameter of 1.5. Fixing the Toomre parameter gives a constraint on both the scale-height and density of the disc as a function of observable spectral parameters in our model.

\subsection{Accretion rates in self-gravitating discs}
In our model, the accretion rate onto the SMS is supplied by turbulent angular momentum transport in the SMD. In full generality, the accretion rate is given by:
\begin{align}
    \dot M_{\rm D} = 2 \pi R_{\rm D} \rho H_{\rm D} v_{\rm r}
\end{align}
where $v_{\rm r}$ is the radial velocity of fluid elements through the disc. We formulate the radial inflow through the standard alpha prescription \citep{Shakura:1973uy} where:
\begin{align}
\label{eq:vr_alpha}
    v_r \sim \alpha c_s h_{\rm D}.
\end{align}
Then, the accretion rate becomes:
\begin{align}
    \dot M \sim  \alpha2 \pi R_{\rm D}^2  \rho h_{\rm D}^2 c_s \ ,
\end{align}
{Where $\alpha < 1$ is required to exclude turbulent eddies larger than a scale-height or supersonic turn-over velocities.} Evaluating for typical parameters gives:
\begin{align}
    \label{eq:accratedisc}
    \dot M_{\rm D} &\approx 9.5 \times \left( \frac{\alpha\, h_{\rm D}^2}{10^{-5}} \cdot \frac{n_{\rm D}}{10^{14}\, {\rm cm}^{-3}} \right) \nonumber \\ &\times \left(\frac{R_{\rm D}}{0.1\, {\rm pc}} \right)^{2} \left( \frac{T_{\rm D}}{4300\, {\rm K}} \right)^{1/2} \, {\rm M}_{\odot}\, {\rm yr}^{-1}.
\end{align}
Inverting Eq. \ref{eq:accratedisc} yields a constraint on the value of $\alpha$:
\begin{align}
     \alpha &\approx 0.1 \times  \frac{\dot M}{10\, M_\odot\, {\rm yr}^{-1}} \left( \frac{h_{\rm D}^2}{10^{-4}} \cdot \frac{n_{\rm D}}{10^{14}\, {\rm cm}^{-3}} \right)^{-1} \nonumber \\ &\times \left(\frac{R_{\rm D}}{0.1\, {\rm pc}} \right)^{-2} \left( \frac{T_{\rm D}}{4300\, {\rm K}} \right)^{-1/2}.
\end{align}
Values of $\alpha \sim 0.1$ are found to adequately fit observed AGN spectra \citep{2007king}. Similarly, in the case of a marginally stable gravito-turbulent disc with a Toomre parameter $\mathcal{S}\lesssim2$, $\alpha$ is also typically found to be of order $\sim 0.1$ \citep{2001gammie,2023chenalpha}.

Given a value of $\alpha$ and a best fit spectral model, equating $\dot M_{\rm S} = \dot M_{\rm D}$ gives a constraint on the disc density and scale-height. {Note that equating the two accretion rates strongly ties the geometry of the disk to the spectral properties of the SMS, resulting in our model being entirely constrained (up to the choice of $f_{\rm c}$) when fit to real spectra. In  our parameter fitting, we} select disc solutions {distributed in a narrow Gaussian around a} nominal value of $\alpha =0.2$, which corresponds to a maximally saturated gravito-turbulent effective viscosity \citep{2023chenalpha}. {We also select Toomre parameters distributed in a narrow Gaussian around a value of $\mathcal{S}=1.5$. Modifying these choice within the constraints discussed above does not influence the results within the uncertainties of the fitting. More details regarding the MCMC procedure can be found in Appendix \ref{app:mcmc}.}

\subsection{Line broadening}
In our model, warm hydrogen lines such as H$_{\alpha}$ and H$_{\beta}$ are emitted by the bulk of the gas in the accretion disc. The circular velocity of the disc is given by:
\begin{align}
    v_{\rm D} &= \sqrt{\frac{G M_{\rm D}}{R_{\rm{D}}}}\nonumber\\
    &\approx 2100\, 
    {\rm km/s}\, \left(\frac{M_{\rm D}}{10^{8}\, {\rm M}_{\odot}} \right)^{1/2}\left(\frac{0.1\, {\rm pc}}{R_{\rm D}} \right)^{1/2}.
\end{align}
The corresponding line broadening at wavelength $\lambda$ is typically described as a Gaussian with a standard deviation proportional to the circular velocity, projected onto the line of sight:
\begin{align}
    \sigma_{\lambda} = \sin(\iota)\frac{v_{\rm D}}{c}\lambda.
\end{align}
Therefore, fitting the broadening of the lines gives us another constraint on the disc parameters, {which links the inclination with the density and gravitational compactness of the disk}. The full set of constraints from line broadening, $\mathcal{S}$, $\alpha$, and the spectral fit completely fixes the disc solution up to different choices of the SMS compression factor $f_{\rm c}$. As we will see in section \ref{S:spectra-fitting}, the disc density $n_{\rm D}$ can be recovered up to a scaling factor $\propto f_{\rm c}^{-2}$ and the aspect ratio $h_{\rm D}$ up to a factor $f_{\rm c}^{2}$. {We discuss the implications of invoking additional line broadening mechanisms, such as electron scattering, in section \ref{S:Dicsussion}.}

% ============================================================
\section{Model fitting to observed spectra}
\label{S:spectra-fitting}
\begin{figure*}
    \centering
    \includegraphics[width=0.49\linewidth]{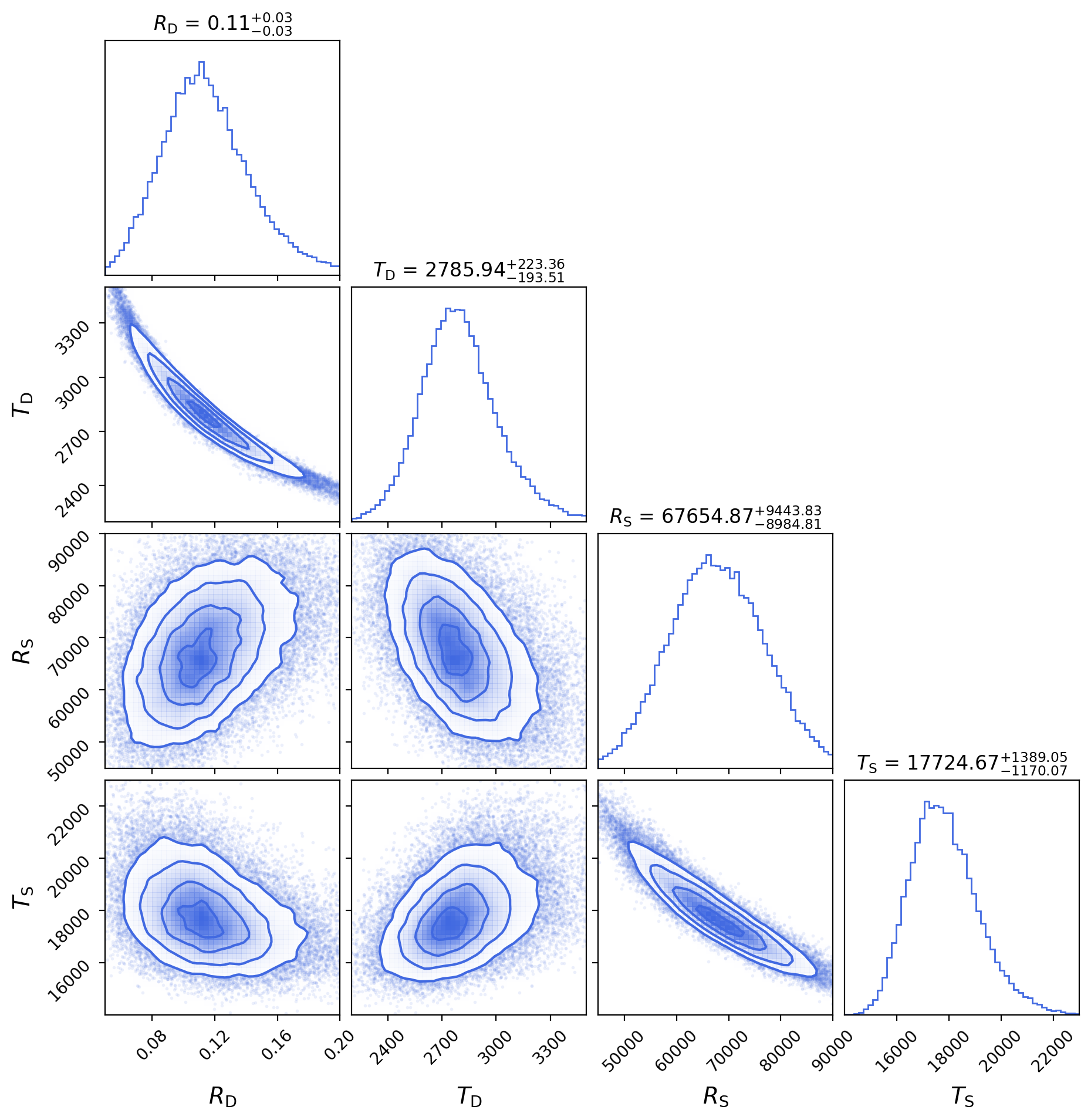}
    \includegraphics[width=0.49\linewidth]{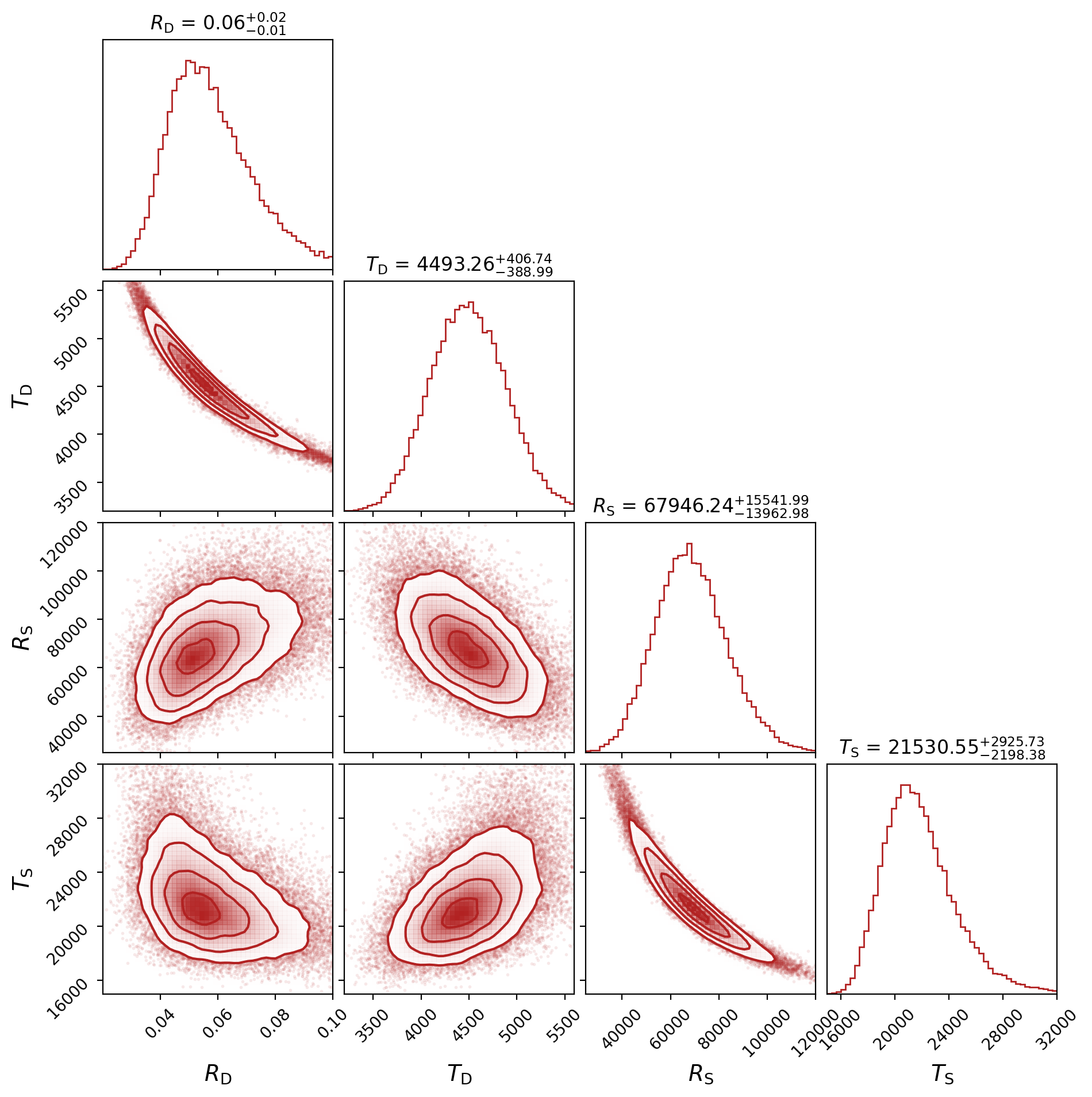}
    \includegraphics[width=0.99\linewidth]{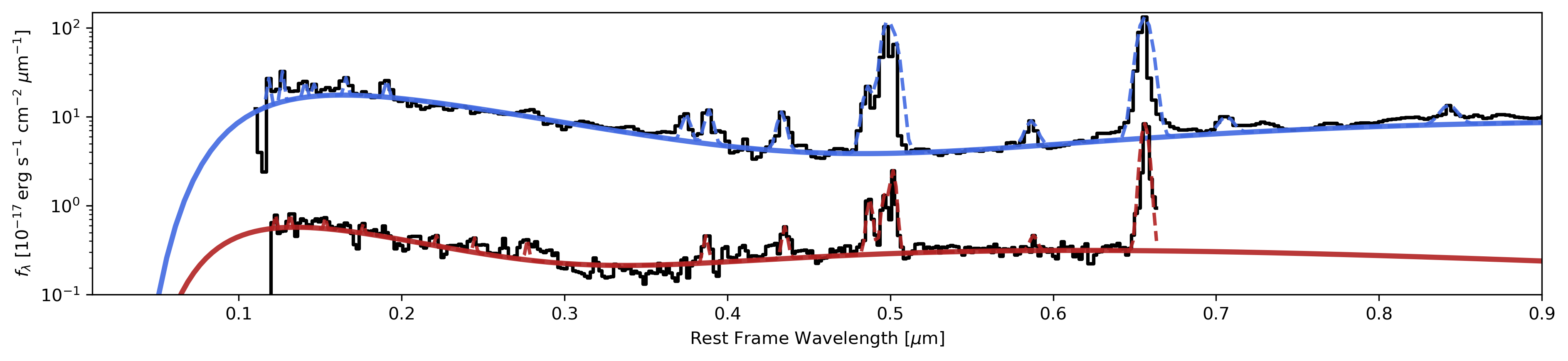}
    \includegraphics[width=0.99\linewidth]{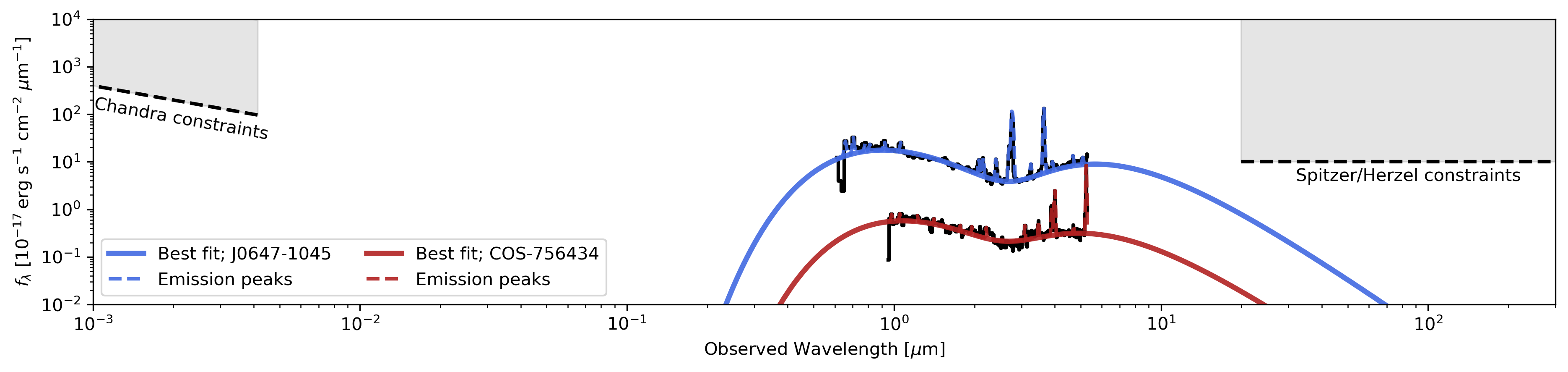}
    \caption{Plot of the best-fit model for J0647-1045 (blue) and COS-756434 (red). The top section shows a part of the recovered posteriors from the MCMC parameter fitting, highlighting a correlation between the SMS/SMD size and temperature. The bottom two panels show the resulting model spectra plotted over the observed data, including and excluding emission peaks. {Note that the height of the emission peaks is not part of the model, and is fitted automatically.} The constraints in the X-ray and IR are adapted from published work \citep{Sachhi:LRD-Xrays:2025, Wang:LRDdust:2025}.
    }
    \label{fig:bestfitspectra}
\end{figure*}

% ------------------------------------------------------------
\subsection{Representative fits for J0647-1045 and COS-756434}
We demonstrate that the SMD/SMS model is able to reproduce observations, while also preserving the physical constraints detailed in section \ref{S:physical-model}, by fitting a small number of spectra. LRD spectra are presented in several publications \citep[e.g.][]{2024setton,Greene:LRDs:2024,Inayoshi:bLRDs:2025}. We select two based on the following criteria: They present with the characteristic spectral V-shape and broad Balmer emission lines, and they do not posses a strong Balmer-break (which is currently not modeled, though see section \ref{S:Dicsussion} for further discussion). Additionally, we prefer LRDs that have been observed at high SNR, with published spectra in plain units of flux.
% \footnote{As opposed to magnitudes in a specific JWST band. Please forgive the theorist prior. \ct{I think this might undermine our credibility a little haha}}
We choose two spectra that follow these criteria denoted as J0647-1045 \citep{2024killi} and COS-756434 \citep{akins2024} located at a redshift of $4.55$ and $6.99$. We perform a least-squares spectral fitting in log-space by employing Monte-Carlo Markhov-Chain (MCMC) methods via the \texttt{emcee} package \citep{2013emcee}, as detailed in Appendix~\ref{app:mcmc} (which also includes the full posteriors for the recovered parameters). The fitting procedure accounts for the physical constraints detailed in sections \ref{S:physical-model}, \ref{sec:sms} and \ref{sec:smd}, as well as the presence of additional absorption and emission which are removed from the spectra to allow for better fitting of the continuum.

The best fit spectra for J0647-1045 and COS-756434 are shown in Fig. \ref{fig:bestfitspectra}, and the best fit parameters that are fully constrained by the MCMC fitting are listed in Table \ref{tab:params}. The 1-d posteriors of these parameters are roughly Gaussian, and we simply report their standard deviation as a measure of the uncertainty on the spectral fit (see Appendix \ref{app:mcmc} for the full posteriors). The SMD/SMD model provides an excellent fit to the observed spectra over the entire observed range, while also following the many physical constraints detailed in section \ref{S:physical-model}. 
Additionally, it respects both higher- and lower-energy observational constraints.
In the bottom panel of Fig.~\ref{fig:bestfitspectra}, we include with our fits upper-limits on the X-ray emission from stacked \textit{Chandra} observations \citep{Sachhi:LRD-Xrays:2025} as well as infrared (IR) constraints from a particularly bright and nearby source which we take as upper-limits \citep{Wang:LRDdust:2025}.
However, we will see that accretion rates below the Eddington limit, and the preservation of a thin disc geometry require values of the compression parameter $f_{\rm c}< 1$, which we will discuss in detail in the next section.

As expected, we recover temperatures of the order of $\sim 4000$ K and $20000$ K for the cold ({disk}) and the hot ({star}) part of the spectrum, respectively. Additionally, the size ratio between the two BB components is of order $\sim 100$, as implied by the relative flatness of the specific flux across the observed band. The required typical size to match the luminosity for the cold part of the spectrum is roughly $\sim 0.1$ pc. We highlight this in particular, as a few works have tentatively associated the redder part of the spectral continuum with a hypothetical structure at the Hayashi limit \citep{Inayoshi:lrdxrays:2024,liuquataert2025}, similar to a SMS. {Crucially, this is the inverse of our interpretation. We can also demonstrate that this identification is inconsistent with constraints on SMS stability: By using Eq. \ref{eq:hosokawa} we can see that a hypothetical SMS following the Hayashi limit reaches sizes of $\sim 0.1$ pc at a corresponding mass of $\sim f_{\rm c}^{-2}\times 3 \times 10^8$ M$_{\odot}$. This is at a minimum two full orders of magnitude larger than the expected onset mass for the GR instability {for non rotating SMS}. Therefore, we deem the association of a "traditional" sub-Eddington SMSs to the cold ($\sim 5000$ K) part of LRD spectra to be incorrect.} Very recently, \cite{2025begelman} suggested that such a large structure may instead be explained by a late-stage quasi-star, i.e. a bloated SMS envelope containing a massive BH, though other spectral features such as the UV peak are not treated in detail. As another possibility, \cite{2025nandal} invoke a SMS accreting at super-Eddington rates and attribute e.g. Balmer break features to details in the stellar atmosphere. While promising in certain aspects, both of these models crucially rely on extreme accretion rates to obtain the desired SMS or quasi-star properties. However, they omit to model the geometry and emission of the accretion flow itself, which should instead be important (by construction). In contrast, our model is closer to \cite{ZhangWu:TwodiskLRD:2025} which associates the cooler part of LRD spectra with the existence of a larger scale self-gravitating disc, while relying on a more standard AGN interpretation for line broadening\footnote{We note that out of the mentioned articles, only \cite{ZhangWu:TwodiskLRD:2025} clearly show well fitting continuum spectral models over the entire observed range.}.

\begin{table}[]
    \centering
    \begin{tabular}{c|c|c}
        & J0647-1045 & COS-756434 \\ \hline
        $z$ &  4.55  & 6.99   \\
        % $T_{\rm S}$ [K] & 17700 $\pm$ 1300 & 21400 $\pm$ 2800   \\
        % $R_{\rm S}$ [R$_{\odot}$] & 67600 $\pm$ 9300 & 68000 $\pm$ 15100  \\
        $T_{\rm S}$ [K] & (1.77 $\pm$ .13) $\times 10^4$ & (2.14 $\pm$ .28) $\times 10^4$   \\
        $R_{\rm S}$ [R$_{\odot}$] & (6.76 $\pm$ .93) $\times 10^4$ & (6.80 $\pm$ 1.51) $\times 10^4$  \\
        $T_{\rm D}$ [K] &2800 $\pm$ 230 &  4500 $\pm$ 400  \\
        $R_{\rm D}$ [pc]& 0.11 $\pm$ 0.03&  0.06 $\pm$ 0.016 \\
        $M_{\rm D}$ [$10^8$ M$_{\odot}$]& 2.0 $\pm$ 1.0 & 0.73 $\pm$ 0.38  \\
        $\iota$ & 19.6$^{ \circ}$ $\pm$ 3.8$^{ \circ}$ & 13.5$^{ \circ}$  $\pm$ 3.1$^{ \circ}$ \\
    
    \end{tabular}
    \caption{Summary of the fully constrained best fit parameters and their uncertainties for the two given LRD spectra.
    }
    \label{tab:params}
\end{table}

% ------------------------------------------------------------
\subsection{{Consistency and results of the model fits for different values of the SMS compression factor.}}
\label{sec:fc}
Several additional physical parameters of J0647-1045 and COS-756434 are constrained up to a choice of the compression factor $f_{\rm c}$. These are reported in Table \ref{tab:paramsfc}. We highlight here some important considerations that are valid for both sources and use these to estimate reasonable values for $f_{\rm c}$.

Crucially, the geometry and density of the SMDs is entirely fixed  given a choice of $f_{\rm c}$, {since both the radius and scale height of the disc are related to the accretion rate which ultimately determines the temperature of the SMS}. As discussed in section \ref{sec:smd}, marginally stable self-gravitating discs in proto-galaxies have typical scale-heights of $h_{\rm D}\sim 0.01$ to $\sim 0.1$, {thanks to strong turbulent support} \citep{2007lodato,Mayer2010,Mayer2015,2019booth}. Adding this constraint suggests values of $f_{\rm c}\lesssim 0.1$. Additionally, it is important to note that no thin disc solutions with $h_{\rm D}<1$ are found for $f_{\rm c}\gtrsim 0.4$ while preserving a Toomre parameter of $\mathcal{S}\sim 1.5$ and a viscosity $\alpha \sim 0.2$ (the precise condition depends on a choice of the latter two parameters). {This arises from the fact that less compressed SMSs require a higher accretion rate to reach a given temperature, which in turn necessitates for a thicker disk once viscosity and density are fixed.} The associated SMD densities are comparable to or lower than typical AGN densities of $\sim 10^{14}$ cm$^{-3}$ \citep{Shakura:1973uy,2002frank,2019jiang} and do not provide further insights (beyond a potential connection to Balmer-breaks discussed in section \ref{S:Dicsussion}).

The recovered masses of the SMSs are comparable for both sources, $\sim (7\times 10^4)\, f_{\rm c}^{-2}$ M$_{\odot}$. Recall that the majority of estimates place the onset of the GR instability for SMSs at $\sim 10^6$ M$_{\odot}$ {for non rotating SMS, and potentially more when additional stabilizing mechanisms are present} (see section \ref{sec:sms}). 
Assuming a constant accretion rate, the timescale for an SMS to change mass considerably is $M_{\rm S}/\dot M$ such that the longest phase of the entire SMS lifetime is just before the onset of the GR instability when the mass doubling timescale is longest. Consequently, the most likely time for an observer to see the system is also close to the maximum SMS mass. We find that a compression factor $f_{\rm c}\lesssim 0.1$ gives a best fit SMS mass on the order of several $10^6$ $M_{\odot}$, {which is plausible for slowly rotating SMS (see again section \ref{sec:sms})}. Crucially, for the recovered parameters, the accretion rate through the disc and onto the SMS scales roughly as $\sim f_{\rm c}^2\times 10^4$ M$_{\odot}$ yr$^{-1}$ (see Eq. \ref{eq:accrate}).

{We can also compare with what is known regarding relaxed SMS. Massive stars are radiation-pressure supported, with a luminosity approaching the Eddington limit \citep{1921eddington,Kippenhahn:1994tm}. Stellar evolution simulations with constant accretion rates (e.g. \citealt{hosokawa2013,haemmerle2018a,2024A&A...689A.351N}) show that after reaching $\sim 10^5$ to $\sim 10^6$ M$_{\odot}$ through accretion, SMS will contract from $R\sim10^4$--$10^5\,R_\odot$ to $R\sim10^2$--$10^3\,R_\odot$, reaching effective temperatures of $T_{\rm S}\sim (1$--$5)\times10^4$ K. \cite{2025arXiv251108516N} show similar qualitative behaviour, though the GR instability occurs at earlier scales due to the presence of metallicity. Thus, these simulations show how SMS transition into a more compact state towards the end of their lifetime, with temperatures appropriate for the hot LRD component and approximately Eddington luminosity. Note here the fact that imposing Eddington limited accretion (see Eq.~\ref{eq:eddi}) on our spectral fitting result similarly, and consistently, requires a compression factor $f_c \lesssim 0.1$, corresponding to accretion rates $\lesssim 10^2$ M$_{\odot}$ yr$^{-1}$.}

{To summarize,} these four separate considerations lead us to postulate a value of $f_{\rm c} \sim 0.1$ {:
\begin{itemize}
    \item Consistency of the recovered parameters with expectations for self-gravitating, marginally stable disks.
    \item  Consistency of the recovered SMS masses with a slightly delayed GR instability.
    \item Consistency with size and temperature expectations in relaxed SMS.
    \item  Consistency of the accretion rates with the Eddington limit.
\end{itemize}}

Remarkably, the same choice for $f_{\rm c}$ will also be sufficient to match the expected galaxy mass -- BH mass relation {and the possible presence of a cutoff luminosity for LRDs} (as detailed in section \ref{S:Dicsussion}).
\newline \newline

Overall, the two sources for which we performed a full MCMC fitting show a remarkable similarity in the recovered best fit parameters. This is partly due to the surprisingly universal features of LRD spectra, and partly due to the extremely weak scaling of the SMS/SMD model with physical parameters (with the exception of the compression parameter $f_{\rm c}$). We claim that the SMS/SMD model is able to adequately fit a large subset of LRD spectra. As seen in Fig. \ref{fig:projections}, slight differences in the spectra can be accounted for by the following: Luminosity and hardness variations in the hot part of the spectra can be attributed to varying the SMS component radius and temperature. Variation in the cold portion of the spectra can occur from varying disc inclination, size and temperature. Additionally, a combination of inclination, disc mass and disc radius can account for variations in the line broadening. Valid solutions that respect the constraints discussed in section \ref{S:physical-model} can always be found, though often require values of the compression parameter $f_{\rm c}\lesssim1$. The only major feature present in a significant subset of LRD spectra that the SMS/SMD model currently does directly address are the prominent Balmer breaks and absorption features seen in a subset of LRDs. These are discussed in more detail in section \ref{S:Dicsussion}.
\begin{table}[]
    \centering
    \begin{tabular}{c|c|c}
        & J0647-1045 & COS-756434 \\ \hline
        $M_{\rm S}$ [$10^4\,$M$_{\odot}$]&   (6.8 $\pm$ 1.9) $\times f_{\rm c}^{-2}$  & (6.8 $\pm$ 3.2) $\times f_{\rm c}^{-2}$   \\
        $\dot M$ [$10^3$ M$_{\odot}$ yr$^{-1}$]&   (25.8 $\pm$ 4.6) $\times f_{\rm c}^{2}$  & (56 $\pm$ 20) $\times f_{\rm c}^{2}$   \\
        $n_{\rm D}$ [$10^{11}$ cm$^{-3}$]&   $\sim 5\times f_{\rm c}^{-2}$  & $\sim  0.6\times f_{\rm c}^{-2}$   \\
        $h_{\rm D}$ &   $\sim 6.0 \times f_{\rm c}^2$  & $\sim 6.0\times f_{\rm c}^2$
    
    \end{tabular}
    \caption{Summary of the additional best fit parameters that depend on $f_{\rm c}$. Note that a value of $f_{\rm c}<1$ is required to preserve a thin disc geometry as well as sub-Eddington accretion rates. The entire SMS/SMD system, including disc geometry and inclination, is determined by spectral fitting up to a choice of $f_{\rm c}$.}
    \label{tab:paramsfc}
\end{table}

% ============================================================
\section{Discussion} \label{S:Dicsussion}

% ------------------------------------------------------------
\subsection{Spectral features}
Our model of a SMS rapidly accreting from a SMD is able to faithfully capture the defining spectral characteristic of LRDs, namely the spectral-V in the rest-frame optical and UV emission.
Beyond this core tenet, however, the model possesses a number of other potentially desirable properties which we discuss qualitatively within the full context of LRD observations.

Of particular note, in the AGN interpretation of LRDs, one of the most puzzling features is the near complete lack of X-ray emission \citep{Yue:LRDs:2024, Sachhi:LRD-Xrays:2025}.
At present these constraints appear to exclude existing appeals to X-ray obscuration \citep{Maiolino:2025}, efficient Compton cooling \citep{MadauHart:2024}, or steep spectral shapes in super-Eddington accreting SMBHs \citep{Inayoshi:lrdxrays:2024, PacucciNarayan:LRDs:2024}.
An accreting SMS, however, is highly extended, and as such would create no relevant X-rays (see Fig. \ref{fig:bestfitspectra}).

Another unique feature of the Balmer emission in LRDs, is that many sources posses strong Balmer breaks, and in some cases break strengths that are too large to explain with an evolved stellar component \citep{NaiduMatthee:2025}, as well as absorption features in the non-resonant H$\alpha$ and H$\beta$ lines.
Because the $n=2$ hydrogen levels that source these features are extremely short-lived, these features require large gas densities \citep{Juodzbalis:2024}.
While the required densities are generally consistent with typical AGN broad line regions (for characteristic ionization temperatures $ \sim 10^4$K), Balmer absorption features are very rare in typical quasars \citep{Aoki:QuasarAbsorb:2006, Hall:QuasarAbsorb:2007, Schulze:QuasarAbsorb:2018, Zhang:QuasarAbsorb:2018}.\footnote{Although we note that Balmer absorption features have been observed in 10-20\% of AGN observed by JWST \citep{Lin:JWSTAGN:2024}.}
This has led many authors to speculate that LRDs are AGN embedded in dense clouds of warm gas with covering factors near unity \citep{Juodzbalis:2024, Inayoshi:Balmers:2025, NaiduMatthee:2025, Kido:LRD:2025}.
For cooler characteristic temperatures ($\sim 5000$K), \citet{liuquataert2025} demonstrated that Balmer breaks (and by necessity, associated absorption features) remain prominent for photospheric gas densities $\lesssim 10^{-9} \unit{g}\, \unit{cm}^{-3}$ with a ``sweet spot'' in the effective continuum opacity ratio\footnote{The ratio of the effective opacity at 4000\r{A} to that at 3600\r{A}, straddling the Balmer break.} in the range $\sim 10^{-10} - 10^{-11} \unit{g}\, \unit{cm}^{-3}$ (see e.g. their Figure 2).
For compression values $f_c \lesssim 0.1$ as determined in Section~\ref{S:spectra-fitting}, the characteristic densities of our SMDs fall precisely in this range (see Table~\ref{tab:paramsfc}), implying that such discs could self-consistently source Balmer breaks and absorption features.
We comment that our model would predict those systems with Balmer features to have comparatively warm discs ($\gtrsim 4000$K) and those without to be colder.
% We comment here that the best fit disc temperature for J0647-1045 ($\sim 3000$K) is likely too cold to source a prominent Balmer break, consistent with its spectrum (while COS-756434 is marginal, and accordingly, ).
However, the particulars of these processes would likely be sensitive to the details of the disc structure and turbulence (and may also emerge from associated accretion winds).
We leave a more in-depth analysis for future work, but note that this intrinsic complexity might account for the variation in the presence, strength, and width of Balmer features, as well as the apparent range of absorption centroids from blue- to red-shifted relative to the line centres \citep{Matthee+2024-LRD, Labbe:LRDbalmerbreak:2024, Kocevski:LRDs:2024}.
% We also note that a self-gravitating SMD naturally produces ``over-broadened'' Balmer lines relative to equivalent virial estimates because of the flatter rotation curve and larger velocities at comparable radius. In tandem with the intrinsic gravito-turbulence, this could account for the non-gaussian wings of the broad lines \citep{Rusakov:Nature:2026}.
%\ct{Maybe we want to also change this last sentence along with our new comments on electron-scattering broadening -- I can do once we've settled on what to say; although this point is also interesting--if you add some random scatter to the Doppler broadening, is it still gaussian? Or is that similar to the random-walk behavior of electron scattering?...}

An originally prominent interpretation of LRDs was highly (or uniquely) dust-obscured AGN, but faint rest-frame near-infrared (NIR) observations suggest a generic lack of hot-dust tori \citep{Eisenstein:2023, Williams:LRDdust:2024, akins2024, Wang:LRDdust:2025, Setton:LRDdust:2025}.
In general, the LRDs tend to show a very flat rest-frame NIR component that is also inconsistent with un-obscured AGN (but may be compatible with a 1.6$\mu$m bump from star-formation; \citealt{Sawicki:2002, perez2024}).
This emission, conversely, is well approximated by a blackbody spectrum with temperature of order $5000K$ \citep{Inayoshi:bLRDs:2025, liuquataert2025, 2025begelman} in qualitative agreement with the Rayleigh-Jeans tail of emission sourced by the SMD in our model.

%Lastly, we comment on the spectral variability expected in the SMS/SMD model. In SMDs, the typical timescale of variability occurs is given by the ratio of the disc's scale-height with the velocity of strong turbulent eddies. Given that the turbulent velocity is $\sim v_{\rm D} h_{\rm D}$, we obtain a typical timescale of:
%\begin{align}
   % \sim 47\, {\rm yr}\, \left(\frac{R_{\rm D}}{0.1\, {\rm pc}}\right)^{3/2}\left(\frac{10^8\, {\rm M}_{\odot}}{M_{\rm D}}\right)^{1/2}.
%\end{align}
%Evaluating for values consistent with the 1$\sigma$ posteriors of COS-756434, we obtain a variability timescale of $\sim 10$ yr. Interestingly, \cite{2025furtak} finds variability of order $10\%$ in the broadening of the Balmer lines in a different LRD than than the ones studied here, on a comparable timescales of $\sim 2.5$ yr. Here we simply we note that strong gravito-turbulence provides a way to explain variations in Balmer line broadening of order $10\%$ that is intrinsic to the model.

\subsection{Rates and redshift evolution}
\label{sec:rates}
The number density of LRDs has been estimated to be of the order $\sim 10^{-5} - 10^{-4}$ Mpc$^{-3}$ \citep{perez2024,Matthee+2024-LRD,2025ma}, with a distribution peaking around redshift $z\sim5$ and distributed across $2 \lesssim z \lesssim 10$ \citep{Kocevski:LRDs:2024}. First we start with a simple order of magnitude estimate for the expected number density of LRDs in our model. We then derive its redshift evolution in more detail. We assume that the parameters recovered from the J0647-1045 and COS-756434 spectral fitting are representative of the entire LRD population as discussed in section \ref{S:spectra-fitting}.

For our fitted LRD spectra, the recovered SMD masses are of the order $\sim 10^8$ M$_{\odot}$, {and we postulate that this mass budget is ultimately sourced by galaxy mergers. As discussed in section \ref{sec:smd}, the formation of SMDs has only been confirmed in the purpose built simulations \citep{Mayer2010,2019Mayer,2024ApJ...961...76M} which involved massive galaxies with $\sim 10^{10}$ M$_{\odot}$ in gas and stars.} On the other hand, the onset of strong inflows comprising upwards of $10\%$ of the galactic gas is a robust and general feature of gas rich mergers \citep{barnes_hernquist,kazantzidis2005}, {and their persistence down to tens of pc is also supported \citep{2010hopkins}}. At $z\sim 5$ and above, the gas mass in galaxies is typically larger than the stellar mass by almost an order of magnitude \citep[see e.g.][]{2018tacconi,2019wiklind, 2022heintz}, {showing that ample budget is available in terms of gas inflows already at much smaller stellar mass.} Therefore, we use the merger rate of galaxies with a total gas mass of $\sim 10^9$ M$_{\odot}$ as a simple proxy for those with sufficient gas to form SMDs with $\sim 10^8$ M$_{\odot}$, {and parametrize the uncertain SMD formation with an efficiency parameter $\epsilon_{\rm SMD}$}. Note that mergers of larger galaxies may also {contribute to the rate, or perhaps} form larger SMDs which could model particularly luminous LRDs. In fact, the distribution of sizes of SMDs can be predicted within the scenario of galaxy mergers, which we leave to future work. 

In the appropriate redshift range, the number density of galaxies with at least $10^8$ M$_{\odot}$ in stars {, which corresponds to roughly $\sim10^{9}$ M$_{\odot}$ in gas,} is \citep{2016Asong,2021mcleod}:
\begin{align}  
    n^{\rm gal}_{8}\approx  10^{-2}\,\,{\rm to}\,10^{-1}\, {\rm Mpc}^{-3}.
\end{align}
Between redshifts $\sim 4 - 8$ (where the majority of LRDs are observed), the rate of major mergers per galaxy is approximately $1$ Gyr$^{-1}$, where we define a major merger to have a mass ratio of $\gtrsim 0.25$ \citep{2009stewart,2015rodrig,2021oleary}. Therefore, approximately every relevant galaxy will have undergone a major merger within this redshift range, and had the chance to form a SMD {with an efficiency of $\epsilon_{\rm SMD}$}.

In our model LRDs are a transient stage that lasts approximately one SMS lifetime $\mathcal{T}_{\rm S}$, i.e. the total amount of time it takes for the star to grow and eventually collapse. We can then estimate the density of observed LRDs to be:
\begin{align}
    \label{eq:lrd_dens}
  n_{\rm LRD} \approx \epsilon_{\rm SMD}\times n^{\rm gal}_{8}\times\frac{\mathcal{T}_{\rm S}}{1\, {\rm Gyr}}, 
\end{align}
where the Gyr in the denominator is approximately the time between $z = 8$ and $z=4$, i.e. an appropriate redshift range. From Eq. \ref{eq:lrd_dens}, we deduce that the observed LRD number density is recovered for typical SMS lifetimes of order $10^6$ yr, {assuming an efficiency of $\epsilon_{\rm SMD}\sim 1$}.

As discussed in section \ref{S:spectra-fitting}, we assume that we are observing the SMS in the same mass decade in which they reach the maximum allowed mass. Using Eq. \ref{eq:accrate}, the SMS lifetime is then approximately given by a mass doubling timescale:
\begin{align}
    \mathcal{T}_{\rm S} &\approx  \frac{M_{\rm S}}{\dot{M}_{\rm S}}\\
    &\approx 10^6 \times \left( \frac{f_{\rm c}}{0.05} \right)^{-4} \left(\frac{T_{\rm S}}{1.7 \times 10^4 \, \rm{K}} \right)^{-4} \left(\frac{R_{\rm S}}{7 \times 10^4\,{\rm R}_{\odot}} \right) \, \rm{yr}
\end{align}
where we inserted the best-fit values of J0647-1045. Note the very strong scaling with $f_{\rm c}$ and the fact that the required value is remarkably close to our estimates $f_{\rm c}\sim 0.1$ based on the accretion disc geometry and the Eddington limit detailed in section \ref{S:spectra-fitting}.

{Within the uncertainties of our estimates, we conclude that if values of $f_{\rm c} \lesssim 0.1$ are physical, the proposed formation pathway for the SMS/SMD model has the chance to adequately reproduce the observed number density of LRDs. However, we stress that this still requires an SMD formation efficiency of $\epsilon_{\rm SMD}\sim 1$, which is most likely an optimistic assumption. Overall, we conclude that recovering the observed abundance of LRDs is challenging in this simple formulation of the merger driven SMD/SMS model. We briefly mention a few ways in which this tension may be reduced, though we leave a more thorough analysis for future considerations. First, the requirements for very massive SMDs may be relaxed if additional line broadening mechanisms are invoked, such as electron scattering \citep{2025begelman,Rusakov:Nature:2026}. Second, sufficient gas inflows may be produced by repeated minor mergers \citep[see][for a similar effect in the local universe]{2014teodoro} in galaxies with significantly larger gas mass budgets. Third, if the putative compacted SMS can intrinsically reach temperatures of $\sim 10^4$ K from internal energy production mechanisms. This would reduce the required accretion rates through the disk and naturally increase the system lifetime. Fourth, the formation of SMD/SMS need not necessarily be caused by mergers, but rather could potentially form in isolation as a precursor to $\sim 10^6$ M$_{\odot}$ direct collapse events in some subset of atomically cooled halos \citep[see][for a distribution of direct collapse BH masses]{LodatoNatarajan:2006}.} Finally, we also comment that the timescale for the SMS collapse after it becomes unstable, i.e. the free fall timescale, is approximately $1 - 10$ yr. Therefore, we would expect approximately 1 in $10^5$ LRD to collapse into AGN while being observed.
% \newline \newline

{While the overall rate is challenging to estimate,} the distribution of LRDs as a function of redshift can be derived in more detail from our model, as it should exactly follow the major merger rate for galaxies with sufficient gas mass. We present here an estimate based on the halo merger rate from the Millennium simulation \citep{2010Fakhouri}:
\begin{align}
    \label{eq:millennium}
    \frac{{\rm d} ^2 \Gamma}{{\rm d} \xi {\rm d} z} = B_1 \left( \frac{M_{\rm{h}}}{10^{12} \, \rm{M}_{\odot}} \right)^{b_1} \xi ^{b_2} \exp \left[\left( \frac{\xi}{B_2} \right)^{b_3}\right] (1+z)^{b_4},
\end{align}
where $\Gamma$ is the total number of mergers for a single dark matter halo of mass $M_{\rm{h}}$, $\xi\leq1$ is the halo merger mass ratio, and the best fit parameters are $[B_1,B_2,b_1,b_2,b_3,b_4] = [0.0104,9.72 \times 9.72,0.133,-1.995,0.263,0.0993]$. The merger rate per co-moving volume is found by multiplying the halo merger rate with the halo mass function. We use the parametrisation proposed by \cite{2008tinker}:
\begin{align}
    \frac{{\rm d}n}{{\rm d} M_{\rm h}} = f(\sigma) \frac{\rho_{m}}{M_{\rm h}} \frac{{\rm d \log \sigma^{-1}}}{{\rm d}M_{\rm h}} ,
\end{align}
where $\sigma^2$ is an average over the density power spectrum, $f$ a function describing the likelihood of a region to random-walk into being over the critical density, and $\rho_{\rm m}$ is the mean density of the universe. We use the best fit parameters reported in \cite{2008tinker} for a value of the over-density $\Delta =200$, which best reproduces a universal result. Note that we assume standard $\Lambda$-CDM cosmology with $\Omega_{\rm m} = 0.27$ and $\Omega_{\rm \Lambda}=0.73$. From this point, the halo merger rate can be related to the galaxy merger rate by a halo mass--galaxy mass relation. We use a fit of the stellar mass to halo mass ratio $\mathscr{F}^{\star}$ \citep{2018MNRAS.477.1822M,2020A&A...634A.135G}:
\begin{align}
    \mathscr{F}^{\star}(M_{\rm{h}}) &= 2D_1(1+z)^{\delta_1}\left[\left(\frac{M_{\rm{h}}}{D_2} \right)^{-\beta} + \left(\frac{M_{\rm{h}}}{D_2} \right)^{\gamma} \right]^{-1},\\
    D_2&= 10^{\delta_2 + z\delta_3}\, {\rm M}_{\odot}, \nonumber\\
    \beta &= z\delta_4 + \delta_5, \nonumber\\
    \gamma &= \delta_6(1+z)^{\delta_7}, \nonumber
\end{align}
where the best fit parameters are $[D_1,\delta_1,\delta_2,\delta_3,\delta_4,\delta_5,\delta_6,\delta_7] = [0.046,-0.38,11.79,0.20,0.043,0.96,0.709,-0.18]$. Finally, we use a simple fit for the gas mass to stellar mass fraction provided in \citep{2018tacconi}, which roughly increases with redshift as $(1+z)^{2.5}$. Collecting all of these ingredients, we obtain a differential merger rate $\mathcal{R}$ of the form:
\begin{align}
    \mathcal{R} \equiv \mathcal{R}(M_{\rm gas/gal},\xi_{\rm gas/gal},z) \, \left[ {\rm Mpc}^{-3}\, {\rm Gyr}^{-1} \right],
\end{align}
which we can evaluate for either a certain gas mass or stellar mass, as well as for a certain mass ratio.

\begin{figure}
    \centering
    \includegraphics[width=0.99\linewidth]{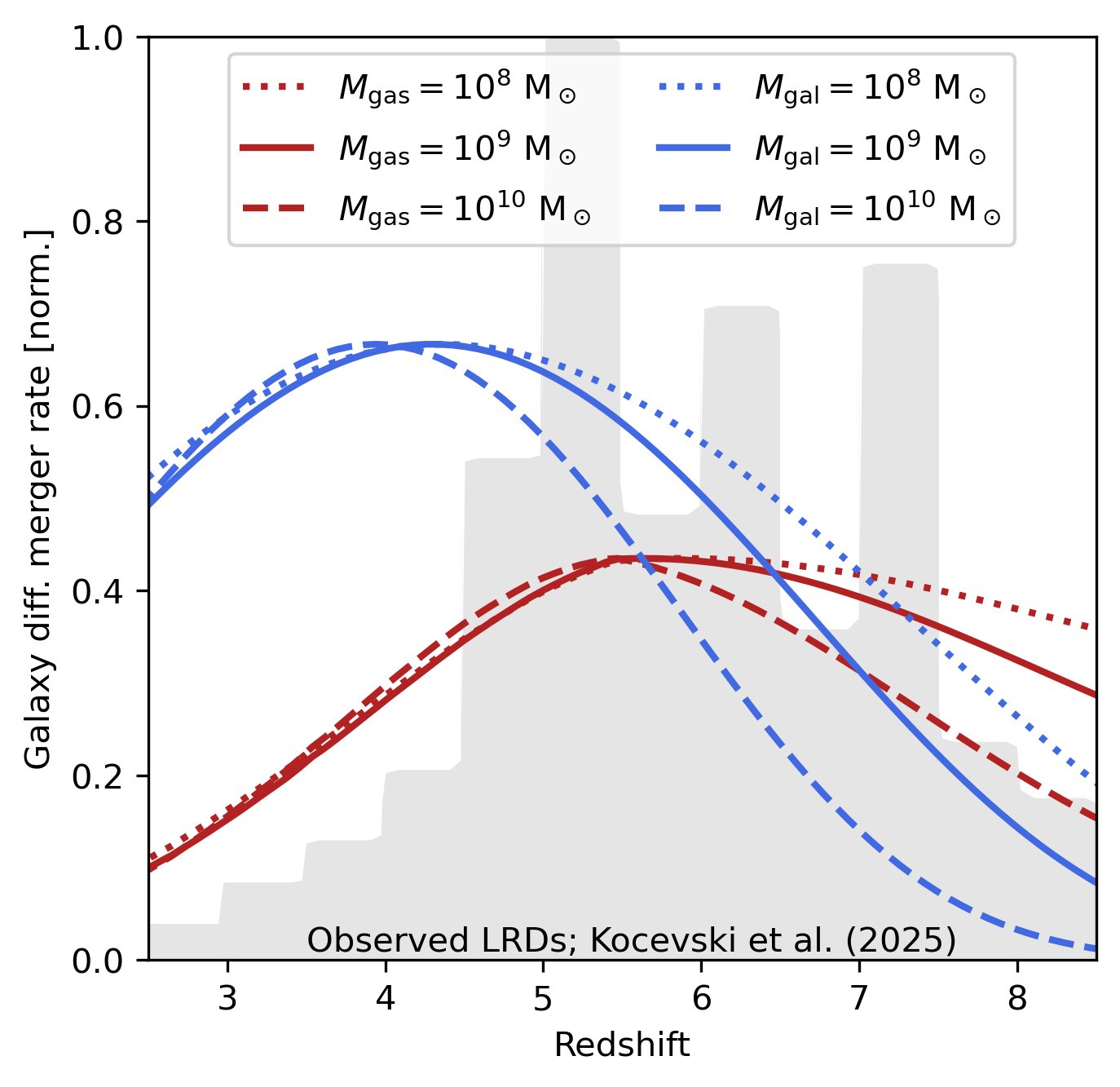}
    \caption{Comparison between the differential merger rates of galaxies with a certain stellar (blue) or gas (red) mass (see text) for a mass ratio of 1.
    The curves are generally normalised to correspond with the high redshift tail for the observed LRD distribution \citep[in gray,][]{2025kocevski}, and different gas (stellar) mass rates are normalized to share the same maximum. Note how the shape of the red curves match the entire observed distribution, strongly suggesting a connection between LRDs and gas rich galaxy mergers with masses $\gtrsim 10^8$ M$_{\odot}$.}
    \label{fig:rate}
\end{figure}

Fig. \ref{fig:rate} shows a comparison of the differential merger rate for galaxies with a stellar (blue) or gas (red) mass of $10^8$ M$_{\odot}$, $10^9$ M$_{\odot}$ and $10^{10}$ M$_{\odot}$. The rates are normalised, and shown for a galaxy mass ratio of 1. 
While we leave a thorough rate calculation as a function of galaxy mass and mass ratio for future work, we note that the distribution in redshift accurately matches the observed LRD distribution for galaxies in the relevant gas mass range. The early peak, at z$\sim 5$, is due to the fact that galactic gas supply is progressively consumed in star formation  such that less gas rich galaxies are available to merge as cosmic time progresses. The preliminary results displayed in Fig. \ref{fig:rate} strongly suggest a connection between gas rich galaxy mergers and the redshift distribution of LRDs, {regardless of the specifics of the SMD/SMS model, and even in a scenario in which matching the overall abundance is challenging.}

% ------------------------------------------------------------
\subsection{Scaling relations and maximum LRD luminosity}
Whereas SMBHs in the local universe typically comprise about $\sim 0.1\%$ of their host galaxy mass \citep{Magorrian:1998cs, ReinesVolonteri:2015}, the estimated masses of SMBHs in LRDs can be as high as $\gtrsim 10 \%$ \citep{Harikane:LRDs:2023, Greene:LRDs:2024, Kocevski:LRDs:2024} with estimates based on assuming the existence of an AGN broad line region. While the intrinsic scatter of galaxy mass--BH mass relations grows with redshift \citep{2010hirschmann}, such over-massive BHs would represent extreme outliers and pose additional challenges for standard galaxy--BH co-evolution \citep{2022habouzit}. The line broadening in our model is set by the size and mass of the SMD rather than a pre-existing BH component, and the BH mass resulting from the  collapse of the central SMS is significantly smaller. Taking a reference value of $f_{\rm c}\sim 0.1$, we recover SMS masses of a few $10^6$ M$_{\odot}$ from our two representative spectral fits (see section \ref{S:spectra-fitting}). Assuming that the majority of the SMS is retained during the collapse, we expect a BH to be formed at comparable mass. The collapse takes place within a recently merged galaxy with at least a stellar mass of several $10^8$ M$_{\odot}$ and a total gas mass of order $10^9$ M$_{\odot}$, where efficient post-merger star formation is taking place \citep{2013hopkins,2015capelo}. Taking these reference values, we see that the predicted BH mass in the SMS/SMD model for LRDs is of order $\sim 0.1 \%$ to $\sim 1 \%$  of its galactic host, landing squarely in the expected range.

Finally, we highlight the recent analysis performed in \cite{Chen:LRDgalaxies:2025}, in which the authors find constraints on the stellar mass of LRD hosts. The recovered constraints broadly imply that typical LRD hosts have stellar masses below few times $ 10^8 - 10^9$ M$_{\odot}$ , where the only constrained galactic component has $10^{8.6}$ M$_{\odot}$. This perfectly matches with the formation pathway discussed in section \ref{sec:rates}. It is particularly intriguing that a majority of the LRDs in the analysis  showcase off-centre emission, which does not have a clear interpretation. 
We speculate that the off-centre emission is a natural result of the galaxy merger, whereby galactic gas clouds collide upon first passage, but the stellar component relaxes over much longer timescales. This is the proposed mechanism of the recently discovered Infinity Galaxy \citep{2025dokkum} in which a direct collapse event is speculated to have occurred.

{Lastly, we comment on the natural and necessary occurence of a typical maximum luminosity for LRDs that is intrinsic to our model. As discussed in section \ref{sec:sms}, the onset of the GR instability is robustly predicted to take place at $\sim 10^6$ M$_{\odot}$ {for highly accreting, non-rotating stars}, and potentially above if if stabilizing forces such as rotation are present. {In section \ref{sec:sms} and across section \ref{S:spectra-fitting}, we have presented several arguments that suggest a value of $\sim 0.1$ for the phenomenological compression factor. Assuming for the moment $f_{\rm c}= 0.1$,} the recovered SMS masses from spectral fitting of $6\times 10^{6}$ M$_{\odot}$, suggest that the observed LRDs consist precisely of such highly accreting SMS {which have been stabilised against the GR instability}, observed in the last mass decade before collapse. {We take the mass scale to be representative, and} assuming accretion at the the Eddington rate, we can use the BB luminosity of the SMS just before the onset of the GR instability to estimate a maximum luminosity:}
\begin{align}
L_{\rm max} &= L_{\rm Edd}(M_{\rm S}^{\rm max}) =\frac{4\pi G M_{\rm S}^{\rm max} m_{\rm p} c}{\sigma_T} \\
    &\approx 6\times10^{44} \left( \frac{M_{\rm S}^{\rm max}}{6 \times 10^6\, {\rm M}_{\odot}} \right)\, {\rm erg/s}.
\end{align}
{While the exact value for maximum SMSs masses and the plausible ranges of the Eddington fraction are unknown, we find that this simple estimate, informed directly from our spectral fits, aligns remarkably well with the cutoff in LRD luminosities reported in \cite{2025ma}.}
% ============================================================
\section{Summary and Conclusions}
\label{S:conclusion}
In this work, we have presented a new physical interpretation of LRDs based on a SMS embedded within a self-gravitating accretion disc (SMD). The SMS/SMD model accounts for a number of key observational features of LRDs, including:
\begin{itemize}
    \item The two-blackbody continuum shape underlying the V-shaped spectra.
    \item The observed broad Balmer emission lines, from the high rotational velocities in the SMD.
    \item The lack of significant X-ray and IR emission, without invoking additional obscuration.
    \item The observed redshift distribution of LRDs, from the distribution of gas-rich major galaxy mergers.
    \item Constraints on the size of LRD galactic hosts.

    \item {The apparent cutoff in number density above luminosities of a few $\times \,  10^{44}$ erg/s}.
\end{itemize}

A key component of the model is the introduction of a phenomenological compression factor $f_{\rm c}$, which parametrises a deviation from  the SMS mass–radius relation of \cite{hosokawa2013}. If a compression (\( f_{\rm c} \lesssim 0.1 \)) is assumed, the model yields long system lifetimes {which are a necessary ingredient} to explain the observed number density of LRDs. Additionally, the model is consistent with sub-Eddington accretion, standard disc geometry, and varied Balmer emission/absorption behaviour. The predicted post-direct-collapse BH masses also land squarely within standard galaxy mass -- BH mass relations, {and the model is also compatible with intriguing features of LRDs such as the presence of off-centre emission}.
All parameters of the SMS/SMD system, including its inclination, can be determined by spectral fitting given a choice of $f_{\rm c}$.

We envision two natural follow-ups this work: First, to thoroughly study the relationship between galaxy merger rates and expected distributions of LRD properties, {including a more extensive series of spectral fits that include Balmer breaks and a quantification of the crucial efficiency parameter $\epsilon_{\rm SMD}$ that governs the formation of SMDs.} Second, to expand the range of simulations of highly accreting SMS in order to more precisely quantify the SMS mass--radius relation of SMS with $M_{\rm S}\gtrsim 10^6$ M$_{\odot}$, beyond the several arguments presented in section \ref{sec:sms}. Such works would strongly establish the plausibility of the SMS/SMD model for LRDs. Together with the arguments presented here, this would bastion the claim that JWST may be directly observing SMS at the verge of relativistic instability, together with their accretion discs: A first direct view of the crucial stage prior to the formation of the original SMBH seeds.

%--------------------------------------------------------------------------
% Acknowledgments%.
%--------------------------------------------------------------------------
\section*{Acknowledgments}
\begin{acknowledgements}
L.Z. is supported by the Villum Fonden grant No. 29466, and by the ERC Starting Grant no. 101043143 -- BlackHoleMergs. L.Z. is also supported by the European Union’s Horizon 2024 research and innovation program under the Marie
Sklodowska-Curie grant agreement No. 101208914, 
and C.T. acknowledges support from the European Union’s Horizon 2023 research and innovation program under Marie Sklodowska-Curie grant agreement No. 101148364.  
L.M. acknowledges support from the Swiss National Science Foundation under the Grant 200020-207406. 
L.Z. and C.T. thank J\'anos Tak\'atsy and Daniel J. D'Orazio for marginally stable discussions. L.M. and C.T. thank Piero Madau, Naoki Yoshida, John Silverman and Zoltan Haiman for useful discussions on LRDs.
\end{acknowledgements}

\newpage
\appendix

\section{Details of the MCMC}
\label{app:mcmc}

\begin{figure}[h!]
    \centering
    \includegraphics[width=1\linewidth]{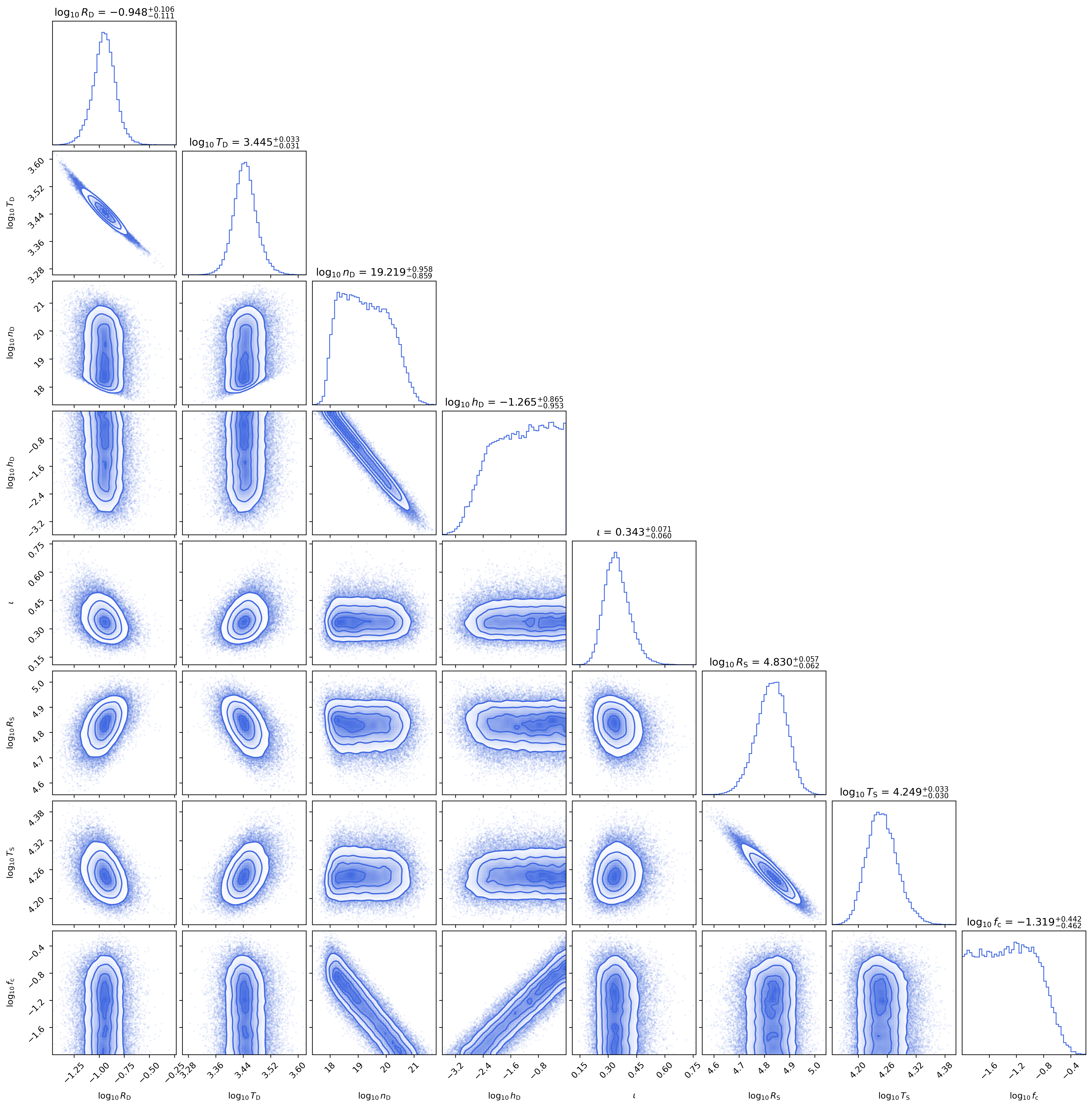}
    \caption{Full posterior for J0647-1045. The units for the parameters [$R_{\rm D},T_{\rm D},n_{\rm D},h_{\rm D}$,$\iota$,$R_{\rm S}$,$T_{\rm S}$,$f_{\rm c}$] are [pc,K,m$^{-3}$,$-$,rad,R$_{\odot}$,K,$-$].}
    \label{fig:full1}
\end{figure}

\begin{figure}[h!]
    \centering
    \includegraphics[width=1\linewidth]{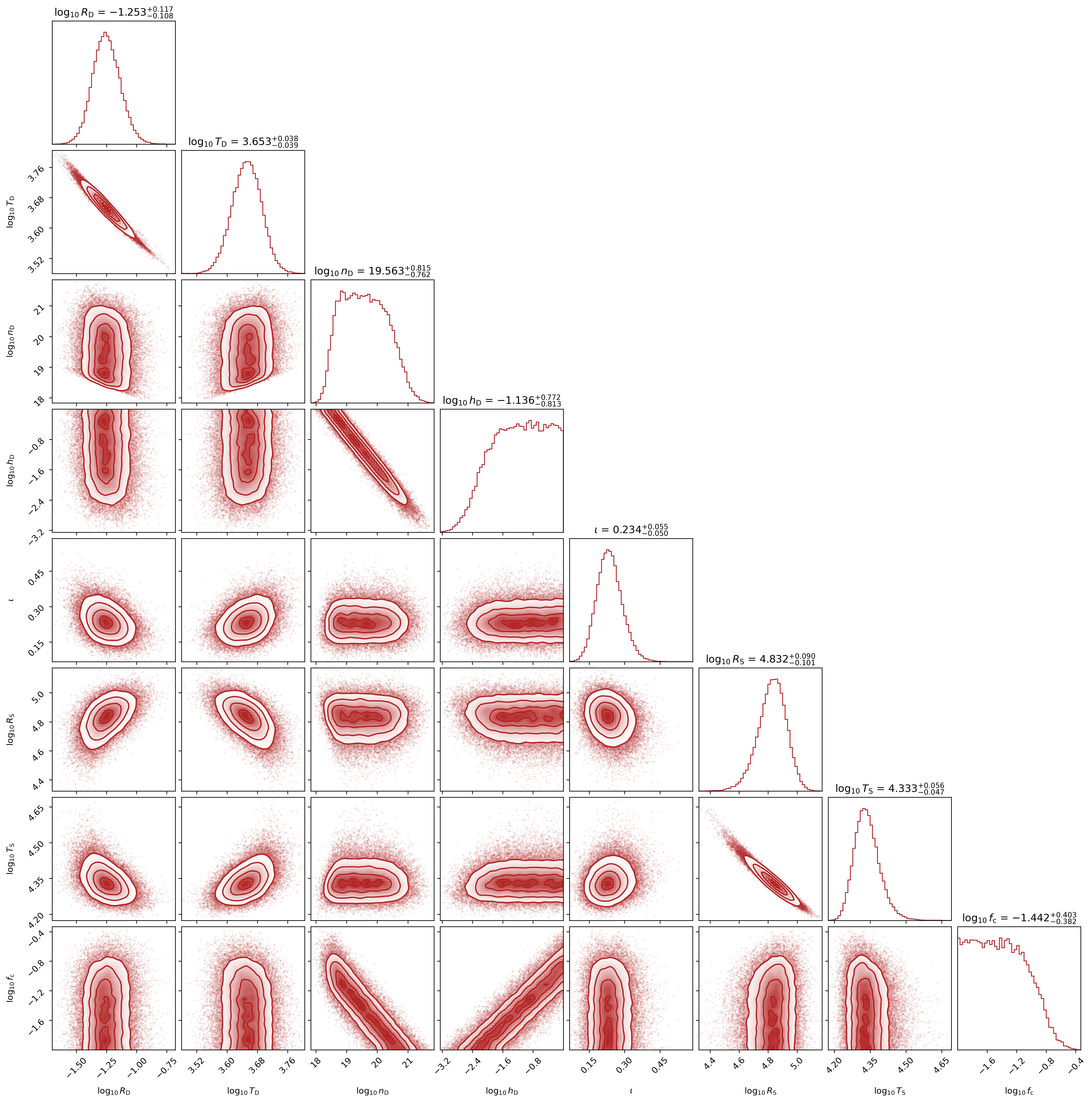}
    \caption{Full posterior for COS-756434. The units for the parameters [$R_{\rm D},T_{\rm D},n_{\rm D},h_{\rm D}$,$\iota$,$R_{\rm S}$,$T_{\rm S}$,$f_{\rm c}$] are [pc,K,m$^{-3}$,$-$,rad,R$_{\odot}$,K,$-$].}
    \label{fig:full2}
\end{figure}

We use Monte-Carlo-Markhov-Chain (MCMC) methods to perform the spectral fitting and estimate parameter correlations numerically. We define a likelihood function $\mathcal{L}$ of the form:
\begin{align}
    \mathcal{L}\left(\Theta \right) \propto \exp\left[ - \big(S(\Theta) - D\big)^2 \right],
\end{align}
where $\Theta = [R_{\rm D},T_{\rm D},n_{\rm D},h_{\rm D},\iota,R_{\rm S},T_{\rm S},f_{\rm c}]$ are the parameters used to evaluate the spectral model $S$. The vector $D$ represents the spectral data of a given LRD, taken with the NIR cam instrument of JWST \citep{2016jdox}, where the total length is given by the spectral range of the measurement divided by the spectral resolution of the instrument (approximately $0.013$ $\mu$m). In practice, we used the data plotted in \cite{2024killi} and \cite{akins2024} for the spectra of J0647-1045 and COS-756434, respectively. Since no errors are given, we treat each data point as being equally weighted and independent.

We use the affine invariant sampler \texttt{emcee} \citep{2013emcee} to perform the numerical tests, running 24 parallel walkers for approximately 20'000 to 30'000 steps. The typical auto-correlation time of the walkers is $\sim 200$. We initialise the walkers by approximately fitting the spectra by eye. The burn-in to find the true maximum of the likelihood required approximately 500 steps and is removed. The priors for the perturbation parameters are initially flat in log-space. The exception is the inclination $\iota$, which has a flat prior ranging from 0 to $\pi/2$. We implement several additional prior constraints on the recovered model parameters in order to enforce the physics discussed in section \ref{S:physical-model}. The prior constraints are as follows:
\begin{itemize}
    \item The SMS maximum accretion rate cannot be exceeded, i.e. $f_{\rm acc}<1$.

    \item The disc must be stable, i.e. $\mathcal{S} > 1$.

    \item The accretion rate must respect turbulent angular momentum transport in the disc, i.e. $\alpha < 1$.
\end{itemize}
Additionally,  we strongly prefer fits that correspond to Toomre parameters of $\mathcal{S}\sim 1.5$ and viscosity prescriptions of $\alpha \sim 0.2$. These are implemented as a Gaussian prior on $\mathcal{S}$ with a standard deviation of 0.25, and a Gaussian prior on $\log_{10}\, \alpha$ with a standard deviation of 0.25. The choices for the standard deviations are arbitrary beyond being numerically convenient and do not strongly influence the uncertainties in the recovered parameters, since these are dominated by other correlations.

In order to accurately fit the basic  SMS/SMD spectral model to the data, we have to account for the various emission and absorption features. Emission and absorption peaks are identified in the data via the \texttt{scipy.findpeaks} function and added automatically to the spectral model as Gaussians. They do not provide any additional constraining power beyond the line broadening, which is fitted exclusively for the H$_{\alpha}$,  H$_{\beta}$ and OIII lines. We show the full posteriors for J0647-1045 and COS-756434 in Figs. \ref{fig:full1} and \ref{fig:full2}, respectively.

\bibliographystyle{aasjournal}
\bibliography{main}

\end{document}